\documentclass[12pt,preprint]{aastex}
\usepackage{natbib}
\begin{document}

\title{Near-Infrared Interferometric, Spectroscopic, and Photometric 
Monitoring of T Tauri Inner Disks} 

\author{J.A. Eisner\altaffilmark{1}, L.A. Hillenbrand\altaffilmark{2},
R.J. White\altaffilmark{3}, J.S. Bloom\altaffilmark{1},
R.L. Akeson\altaffilmark{4}, C.H. Blake\altaffilmark{5}}
\email{jae@astro.berkeley.edu}
\altaffiltext{1}{University of California at Berkeley, 
Department of Astronomy, 601 Campbell Hall, Berkeley, CA 94720}
\altaffiltext{2}{California Institute of Technology,
Department of Astronomy MC 105-24,
Pasadena, CA 91125}
\altaffiltext{3}{University of Alabama in Huntsville, Department of Physics,
201B Optics Bldg, John Wright Drive, Huntsville, AL 35899}
\altaffiltext{4}{California Institute of Technology, 
Michelson Science Center MC 100-22,
Pasadena, CA 91125}
\altaffiltext{5}{Harvard College Observatory, Cambridge, MA 02138}
\keywords{stars:pre-main sequence---stars:circumstellar 
matter---stars:individual(CI Tau, DK Tau, AA Tau, RW Aur, V1002 Sco,
V1331 Cyg, DI Cep, BM And)---techniques:high angular 
resolution---techniques:interferometric}

\begin{abstract}
We present high angular resolution observations with the Keck Interferometer,
high dispersion spectroscopic observations with Keck/NIRSPEC, and near-IR
photometric observations from PAIRITEL of a sample
of 11 solar-type T Tauri stars in 9 systems.  
We use these observations to probe the 
circumstellar material within 1 AU of these young stars, measuring the
circumstellar-to-stellar flux ratios and angular size scales of the
2.2 $\mu$m emission.  Our sample spans a range of stellar luminosities and
mass accretion rates, allowing investigation of potential correlations 
between inner disk properties and stellar or accretion properties.
We suggest that the mechanism by which the dusty inner disk is truncated
may depend on the accretion rate of the source; in objects with low
accretion rates, the stellar magnetospheres may truncate the disks, while
sublimation may truncate dusty disks around sources with higher 
accretion rates.
We have also included in our sample objects that are known to be
highly variable (based on previous photometric and spectroscopic observations),
and for several sources, we obtained multiple epochs of spectroscopic and 
interferometric data, supplemented by near-IR photometric 
monitoring, to search for inner disk variability.
While time-variable veilings and accretion rates
are observed in some sources, no strong evidence for inner disk pulsation
is found.
\end{abstract}

\section{Introduction}
Protoplanetary disks are an integral part of the star and planet formation
process, and studying such disks around early analogs of stars like 
our own sun can provide insight into how planetary systems form.
T Tauri stars are $\la 2$ M$_{\odot}$ pre-main sequence objects, and their
circumstellar environments are an ideal laboratory for studying the initial
conditions of planet formation.  A wealth of evidence, including 
direct imaging at millimeter and optical wavelengths 
\citep[e.g.,][]{KS95,DUTREY+96,MO96}, and modeling of spectral energy 
distributions \citep[SEDs; e.g.,][]{ALS88,BBB88,BECKWITH+90}, has confirmed 
the long-espoused hypothesis that T Tauri stars are surrounded by substantial
disks of dust and gas.  Moreover, observed line profiles and UV continuum
excesses indicate that T Tauri stars are 
accreting material from their circumstellar disks \citep[e.g.,][]
{WALKER72,BB89,EDWARDS+94,GULLBRING+98}.

Study of the innermost disk regions may reveal the mechanism by which material 
is accreted through the disk onto the star.  In the current paradigm, T Tauri 
disks are truncated by the stellar magnetosphere
within the co-rotation radius, with material accreting along magnetic
field lines onto high-latitude regions of the star 
\citep[e.g.,][]{KONIGL91,SHU+94}.  For typical T Tauri star masses,
radii, magnetic field strengths, and accretion rates, predicted truncation 
radii range from $\sim 0.02-0.2$ AU \citep[e.g.,][]{JOHNS-KRULL07}.  
Previous studies have shown that 
the dusty component of disks is generally truncated further out than the
magnetospheric radius \citep[e.g.,][]{EISNER+05,AKESON+05}.
Dust sublimation is caused by heating 
from the central star, plus possible additional heating due to viscous
dissipation of accreting material and luminosity from the
impact of the accretion flow onto the stellar photosphere.
If the accretion luminosity varies with time, then
the inner dust disk edge may pulsate as the sublimation radius moves in or out.
Some evidence for such variability may be present in near-infrared photometric
monitoring data \citep[e.g.,][]{SKRUTSKIE+96,CHS01,EIROA+02}.
Below, we discuss a search for inner disk pulsations using multiple epochs of 
spatially resolved observations at $2.2$ $\mu$m.

Knowledge of the spatial and temperature structure of protoplanetary disks at 
sub-AU radii is important for understanding the properties of dust and gas in 
regions where terrestrial planet formation occurs.
Temperature is the crucial parameter in determining the
radii where dust and water ice, key building blocks of terrestrial planets, 
can condense \citep[e.g.,][]{HAYASHI81}.
Spatial structure is important for understanding how
the close-in extra-solar planets discovered by radial velocity surveys
\citep[e.g.,][]{MARCY+05} either formed at, or migrated to, their observed
orbital radii \citep[e.g.,][]{LBR96}.

Infrared interferometric observations are necessary to spatially resolve
disk regions within $\sim 1$ AU of the central stars.  Such spatially resolved
observations are the only way to probe directly the geometry and 
temperature of these inner disk regions.  The Keck Interferometer has recently 
enabled spatially resolved observations of protoplanetary disks around several 
approximately solar-type T Tauri stars 
\citep{COLAVITA+03,EISNER+05,ECH06,AKESON+05b};
these observations constrained the dust disk inner radius and temperature,
and showed evidence that the inner disk gas extends further in toward the
central star.  Moreover, when these data are combined with previous 
spatially resolved observations of more massive, luminous objects
\citep{MILLAN+99,MST01,EISNER+03,EISNER+04,EISNER+07a,LEINERT+04,
VANBOEKEL+04,MONNIER+05,MONNIER+06,AKESON+05,LIU+07}, the
inner disk properties as a function of luminosity can be investigated
\citep[e.g.,][]{MILLAN-GABET+07}.  In general, these observations 
are consistent with expectations for disks with directly illuminated inner 
walls; however, there are notable exceptions for solar and sub-solar mass T 
Tauri stars, motivating further observations. 

Here we present near-IR interferometric observations of a sample of low-mass
(including solar and sub-solar mass)
T Tauri stars.  These observations spatially resolve the near-IR emission
and allow measurements of the size of the emitting region.  In addition, we 
use spectroscopic and photometric measurements to constrain the relative
stellar and circumstellar contributions to the near-IR emission, as well
as the accretion luminosities of our targets.   
All of these data constrain the 
inner disk properties of our sample, and in cases where multiple epochs of 
data were obtained, enable study of inner disk variability.

\section{Observations \label{sec:obs}}

\subsection{Keck Interferometer Observations \label{sec:ki}}
We used the Keck Interferometer (KI) to obtain spatially resolved measurements
of the near-IR emission from a sample of 11 T Tauri stars (Table 
\ref{tab:sample}).
In addition, we incorporate into our dataset a measurement of RW Aur A
obtained on 23 October 2002 by \citet{AKESON+05b}.  Among the objects for which
we have multiple epochs of data, DK Tau A, AA Tau, and RW Aur A are known
to be photometrically variable at optical through near-IR wavelengths
\citep{HERBST+94,SKRUTSKIE+96,EIROA+02}.

KI is a fringe-tracking long baseline near-IR Michelson
interferometer that combines the light from the two 10-m Keck apertures
over an 85-m baseline \citep{CW03,COLAVITA+03}.  
For each target, we measured squared visibilities ($V^2$) at $K$-band
($\lambda_0=2.2$ $\mu$m, $\Delta \lambda = 0.4$ $\mu$m).  The system
visibility (i.e., the point source response of the interferometer)
was measured using observations of unresolved calibrators 
(Table \ref{tab:sample}),
weighted by the internal scatter in the calibrator and the temporal 
and angular proximity to the target source \citep{BODEN+98}. 
Source and calibrator data were corrected for 
detection biases as described by \citet{COLAVITA99} and averaged into 5s 
blocks. The calibrated $V^2$ for a target source is the average of the 5s 
blocks in each integration, with uncertainties given by the quadrature addition
of the internal scatter and the uncertainty in the system visibility. 
Systematic $V^2$ effects associated with enhanced performance of the
AO system on one telescope relative to the other (which may depend on the
brightness of the target) are
calibrated by applying a ``ratio correction'' \citep[e.g.,][]{COLAVITA99}.
Typical uncertainties are $\sim 5\%$.

\begin{figure}
\plotone{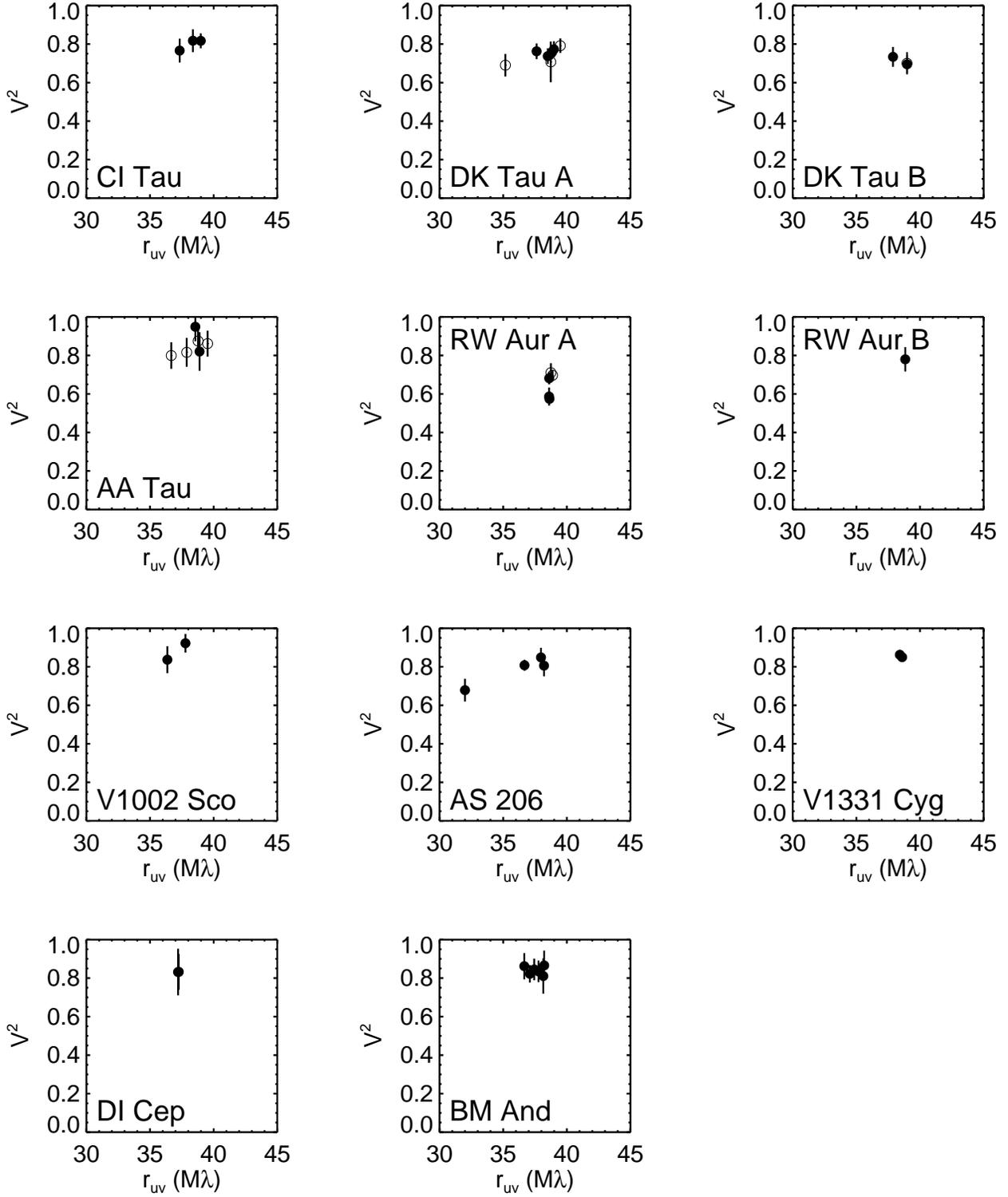}
\caption{Measured $V^2$ from KI as a function of $uv$ radius, $r_{\rm uv}$
($r_{\rm uv}$ is equal to the projected baseline length divided by
the wavelength).
Given the angular resolution of our
observations, these measurements include contributions from the unresolved
central star and resolved circumstellar emission.  
Different epochs are indicated
with different symbols: unfilled circles represent the second epoch
(where available).    The first epoch of KI data for RW Aur A are from
\citet{AKESON+05b}.
\label{fig:raw}}
\end{figure}

Measured $V^2$ are shown in Figure \ref{fig:raw}. 
The angular resolution of our observations is on the order of a milli-arcsecond
(mas), which is sufficient to resolve the near-IR circumstellar emission, but 
not the central star.  For expected stellar radii of 2--5 $R_{\odot}$ and 
distances of 140--700 pc for our sample, the stars subtend $<0.2$ mas . 
Our measured $V^2$ thus include contributions from both
the unresolved stellar component and the resolved circumstellar component.


\subsection{Keck/NIRSPEC Observations \label{sec:nirspec}}
We obtained high-dispersion $K$-band spectra with NIRSPEC between November 2004
and July 2006 for our sample (Table \ref{tab:sample}). 
We used the high-dispersion mode with the 3-pixel slit, 
which provides a resolving power of $R \sim 24,000$.
The NIRSPEC-7 filter was used with an echelle position of 62.67 and
a cross disperser position of 35.51.  This provided seven spectral orders
covering portions of the wavelength range between 1.99 and 2.39 $\mu$m.  
Included in these orders are various stellar atomic lines, the CO (2-0), (3-1),
(4-2), and (5-3) bandheads, and the Br $\gamma$ line.

The spectra were calibrated and extracted using the ``REDSPEC'' package
\citep[e.g.,][]{MCLEAN+03}.  Reduction included mapping of spatial distortions,
spectral extraction, wavelength calibration, heliocentric radial velocity 
corrections, bias correction, flat fielding, and sky subtraction.
We divided our target spectra by A0V or A1V stellar spectra (that were
interpolated over broad Br$\gamma$ absorption features) and 
multiplied by appropriate blackbody templates to calibrate the bandpass of the 
instrument and correct for telluric lines. For the two binary systems 
observed, which have separations of $2\rlap{.}''3$ (DK Tau AB) and
$1\rlap{.}''4$ (RW Aur AB),  the components are sufficiently well separated 
that their spectra could be extracted independently.


\subsection{PAIRITEL $JHK_{\rm s}$ Photometry \label{sec:pairitel}}
We observed our sample with the 1.3-m 
Peters Automated InfraRed Imaging Telescope \citep[PAIRITEL;][]{BLOOM+06}, 
and measured photometric fluxes at $J$, $H$, and $K_{\rm s}$ bands.
For each source, we obtained short exposures (52 ms) and relatively
longer exposures (7.8 s).  Short exposures were read out as ``clearing
exposures'' before the longer exposures (this is the nominal process of
double-correlated reads in infrared imaging).  The short exposures provide 
unsaturated images of our targets, while the longer exposures yield high 
signal-to-noise images of fainter stars within the observed fields
that are used for photometric 
calibration.  The camera provides $2''$ pixels over a 
$8\rlap{.}'5 \times 8\rlap{.}'5$ field of view.  Images are obtained
simultaneously in $J$, $H$, and $K_{\rm s}$ as part of the camera design.

Our data reduction consisted of background (sky+dark) 
subtraction (using dithered
images to produce backgrounds) and flat fielding.  Counts for target and
calibrator objects were measured in apertures with 2-pixel radii, sky 
background was measured as the median value of pixels in an annulus spanning 10
to 20 pixels in radius, and the sky was subtracted from each pixel in the 
aperture. We calibrated the fluxes using measured 2MASS magnitudes for 
other stars within the field of view of each target image.  

Our photometric errors include scatter in measurements
of source and calibrator counts, scatter in the magnitudes computed using
different calibrators, and an assumed uncertainty in the 2MASS magnitudes
of our calibrator stars of 0.017 mag.  Measured fluxes and uncertainties
for our sample are listed in Table \ref{tab:phot}. 

Where possible, we selected calibrator stars that did not appear to vary from 
epoch to epoch; since no photometric standard stars were observed, our
photometry is measured relative to calibrator stars within the observed fields.
We evaluated calibrator variability by comparison of each
calibrator to the mean value of all calibrators within an epoch; as we 
discovered variable sources, we excluded them from the calculated mean value
and iterated.  However, for some fields where only a few stars were visible,
no non-variable calibrator stars were found with this procedure (which may
indicate unstable atmospheric or instrumental conditions); for these 
cases our photometric uncertainties include the apparent variability of the 
flux calibrators.  

The binaries in our sample (DK Tau AB and RW Aur AB) are not resolved with the
$2''$ pixels of the PAIRITEL camera.  Thus the photometry for these objects
is the composite of both components.  The $K$-band
brightness ratios for DK Tau AB and RW Aur A are 3.3 and 4.3, respectively
\citep{SIMON+95,WG01}, and our photometry is thus dominated
by the primary components.

\section{Analysis}
In this section we use the data described in \S \ref{sec:obs} to constrain
the properties of the inner disk regions around our sample objects.
Our focus is on spatially resolving the circumstellar emission with
our interferometry data.  However, because we require measurements of the
circumstellar-to-stellar flux ratio in order to properly model the 
circumstellar component of our $V^2$ measurements \citep[e.g.,][]{EISNER+05},
we begin by deriving this ratio from our NIRSPEC
data (\S \ref{sec:rk}).  In \S \ref{sec:rin}, we model our spatially
resolved interferometric measurements of the circumstellar emission from our
targets with a simple geometric model.  In \S \ref{sec:mdot},  we derive
accretion luminosities for our targets, to enable an investigation of how
the inner disk geometries inferred from our interferometry data correlate
with mass accretion rate.

\subsection{Circumstellar-to-Stellar Flux Ratios \label{sec:rk}}
We use the NIRSPEC measurements described in \S \ref{sec:nirspec} to derive
2 $\mu$m veilings for our sample.  The veiling is defined as
the ratio of the circumstellar excess flux to the stellar flux in the $K$-band.
Veilings are derived by comparing spectra of our target objects with 
spectra obtained (on the same night) for non-accreting, weak-lined T Tauri 
stars (WLTTs). These WLTTs provide excellent templates, since they have 
stellar effective temperatures and surface gravities similar to the classical 
T Tauri stars in our sample, but show neither near-IR excess emission
nor accretion signatures.

We select several atomic absorption lines for comparison between targets and
templates. 
For a grid of WLTT templates, we perform a $\chi^2$ minimization between
the lines seen in the target star and the same lines in the template stars
after registration of the relative radial velocities,
rotational broadening, and veiling.  Radial velocity registration is 
accomplished via a cross-correlation of target and template star spectra.  
Rotational broadening is implemented by convolving the template spectrum with 
a Doppler line profile \citep{GRAY92}.  The template is veiled by adding a
continuum excess to the normalized, un-veiled spectrum, and then re-normalizing
the sum \citep[e.g.,][]{BB90}:
\begin{equation}
F_{\nu, \rm veiled} = \frac{F_{\nu} + r_K}{1 + r_K}.
\label{eq:rk}
\end{equation}

For objects observed over multiple epochs, we first fitted $r_K$ and $v\sin i$
simultaneously for each epoch, then fixed the $v \sin i$
as the weighted average of values from both epochs (which were consistent 
within uncertainties in all cases), and then repeated the fits with $v\sin i$ 
fixed to this average value.  Most of the fitted $v\sin i$ values are smaller
than the velocity resolution of our observations, $\sim 12$ km s$^{-1}$.
This means that the $v\sin i$ values of these targets are compatible
with those of the WLTT templates.  Since
the WLTTs used as templates in our veiling determinations may not be
slow rotators, their spectra may already be rotationally broadened.
Inferred values of $v\sin i$ for our targets thus represent only the
amount by which the template spectra need to be further broadened, rather
than the true $v\sin i$ of the source, and are
only lower limits on the true $v \sin i$ of our sample. 

\begin{figure}
\plottwo{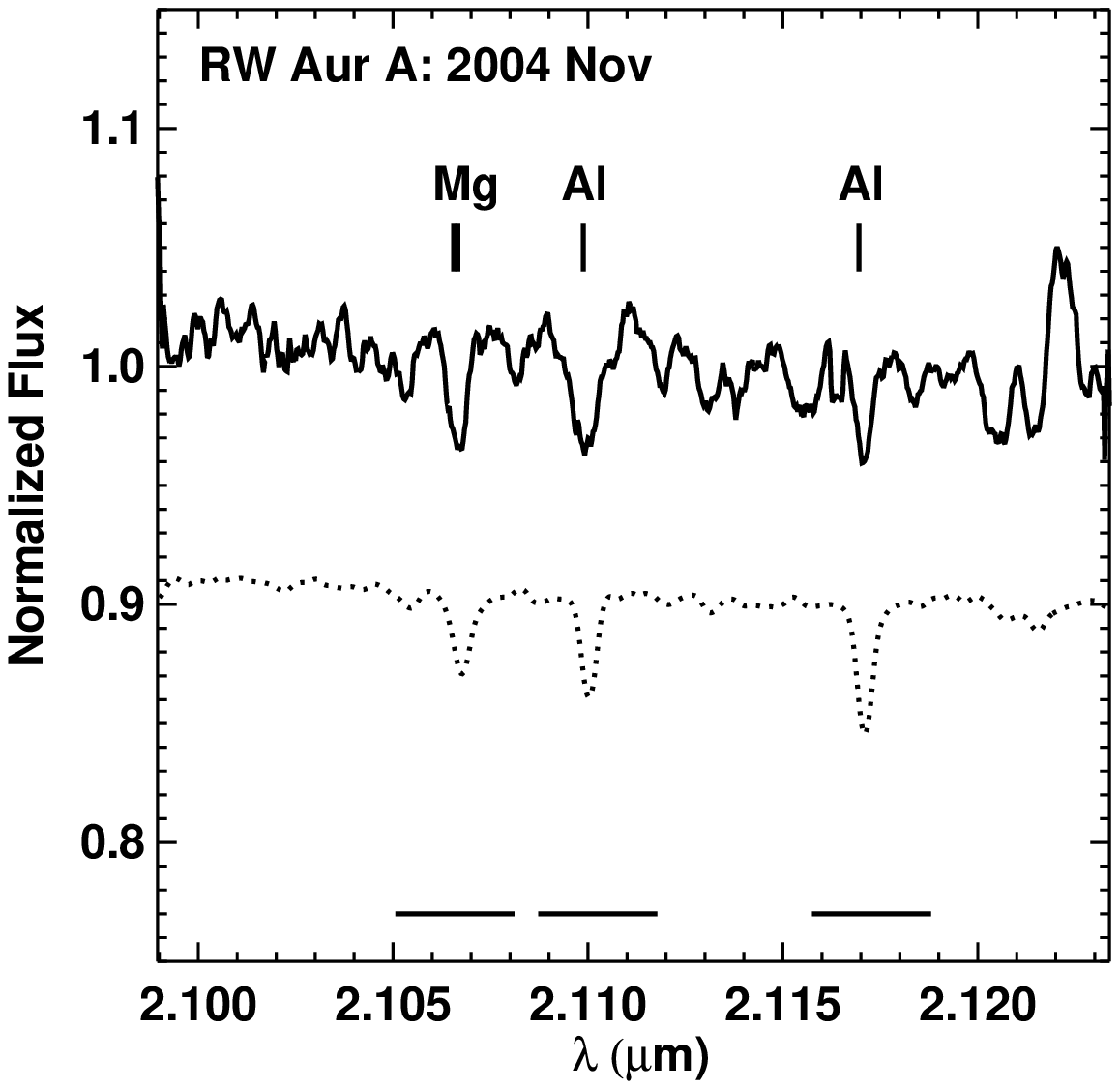}{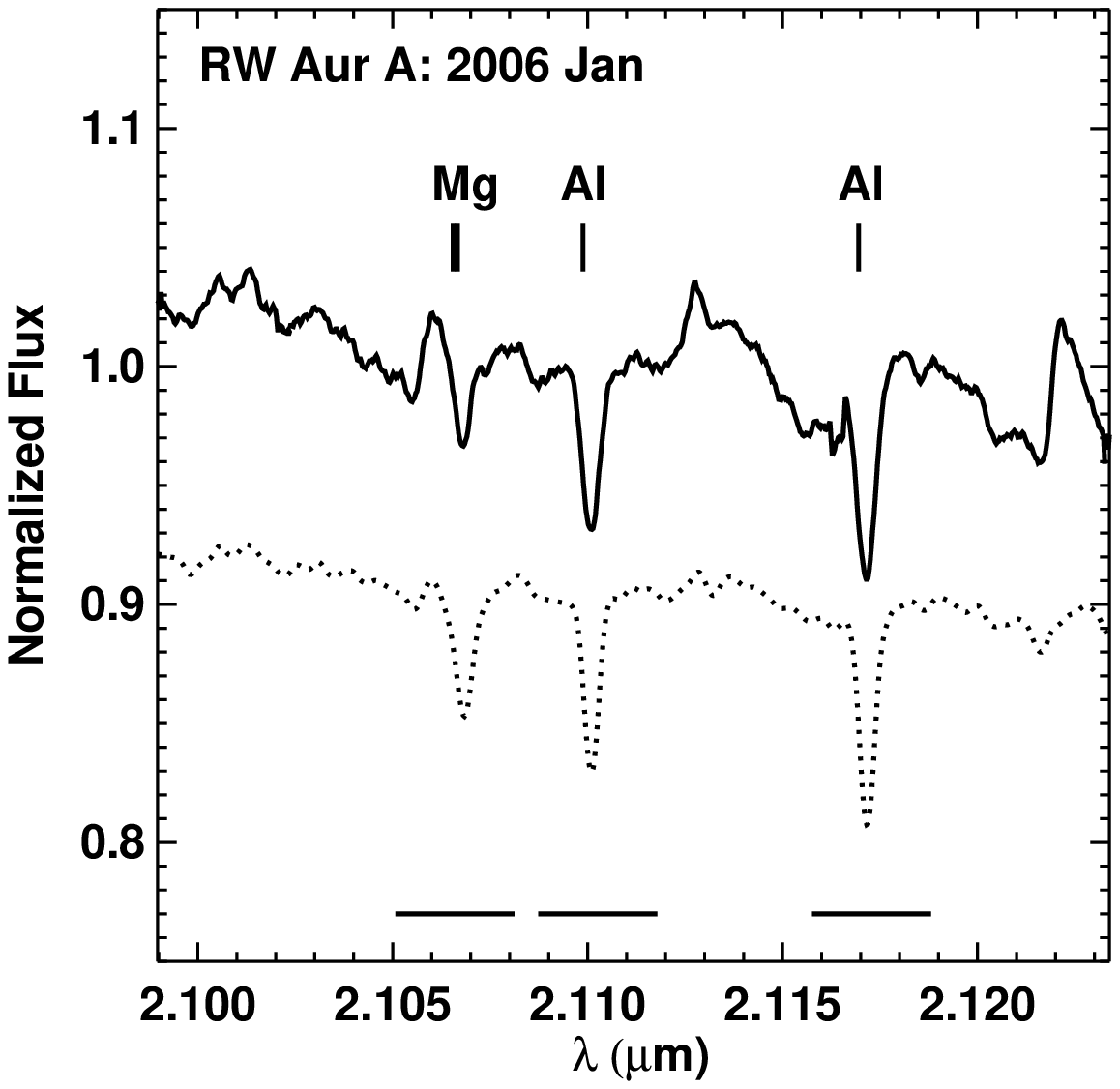}
\caption{Portions of NIRSPEC spectra for RW Aur A ({\it solid curves}), 
with the wavelengths of photospheric Mg and Al absorption lines 
\citep[e.g.,][]{PGS03} indicated.  The spectrum of a non-accreting WLTT 
(AG Tri A), which has been rotationally broadened and veiled
(and vertically offset from the target spectrum), is also plotted 
({\it dotted line}). Spectral regions used when matching rotationally
broadened, veiled template spectra to target spectra are indicated by
straight horizonal lines.
In 2004 Nov ({\it left}), photospheric absorption lines 
from Mg and Al were significantly more veiled than in 2006 Jan 
({\it right}).  The
inferred veilings are $r_K=3.26$ and 1.50, respectively.  
We note the presence of emission from the H$_2$ line at 2.12182 $\mu$m in both 
epochs;  however analysis of this feature is beyond the scope of this work.
\label{fig:nirspec}}
\end{figure}

For this fitting, we used Mg and Al
lines between 2.10 and 2.12 $\mu$m.  
We restricted the fits to 30 \AA \ regions around strong
lines (Figure \ref{fig:nirspec}) to avoid biasing the fits with
noisy continuum regions of the spectra.  As a check, we applied this fitting
procedure to lines of Ti between 2.22 and 2.24 $\mu$m, 
in a different spectral order than the Mg and Al lines; the fitted veilings
in both spectral orders agree within $1\sigma$ uncertainties. 

When we fitted veiling to any of our WLTT templates, using the set of remaining
WLTTs as templates for the fitting, we found veilings consistent with zero 
for all objects but one.  For TYC 5882,
the template with the earliest spectral type, the fitted veiling is
$0.2 \pm 0.1$; this is probably due to the mismatch between the spectral
type of this and other templates (see Table \ref{tab:sample}), 
and does not indicate that TYC 5882 is actually veiled.

There may be some additional uncertainty in the inferred veilings for 
V1331 Cyg and DI Cep, since these sources
have earlier spectral types (G5 and G8, respectively)
than any of our templates (M2--K3; Table 1).  
To evaluate the effects of this source/template mismatch, we
fitted our veiled, rotationally broadened templates to a non-accreting, 
$\sim 30$ Myr old G3/5 star \citep[HIP 9141;][]{ZS04}. 
We found that the best-fit veiling for
this star is $\sim 0.5$, and we thus the veilings inferred for V1331 Cyg and
DI Cep may be biased
by this amount.  We conservatively treat this bias as an uncertainty,
added in quadrature to the statistical uncertainties in the fitted veilings; 
this additional uncertainty is included in Table \ref{tab:veilings}.

Our veiling measurements fall within $\sim 1$ month of our KI observations,
providing estimates of the circumstellar-to-stellar flux ratios nearly
contemporaneous with our interferometric measurements.  Moreover, this
near-coincidence allows a comparison of veiling and the size scale 
of the circumstellar emission at multiple epochs (although there may be
some variability in both quantities over the $\sim 1$ month timescales
separating the spectroscopic and interferometric observations; see below). 
Unfortunately, no measurement
of the $K$-band veiling is available contemporaneous with the 2002 KI 
measurements of \citet{AKESON+05b}.  These authors estimated a 
circumstellar-to-stellar flux ratio of 0.6 by assuming that optical photometry 
traced the stellar photosphere, extrapolating the photosphere to $K$-band
and then comparing the extrapolation to the measured $K$-band flux.  Since the 
star may be veiled at optical wavelengths (due to emission from an accretion 
shock),  the true photospheric flux may be lower than assumed, and their
estimate of the $K$-band veiling is therefore a lower limit.  Nevertheless,
we adopt the value of 0.6 here, and assume an uncertainty of 20\%.

Inferred veiling  values are 
listed in Table \ref{tab:veilings}.  For AA Tau and RW Aur A, 
the veiling varies (at greater than the estimated 1$\sigma$
uncertainty level) from epoch to epoch.  This indicates that the relative
fluxes of the stellar and circumstellar components in these systems are
changing with time.

\subsection{Inner Disk Sizes \label{sec:rin}}
We now turn to analysis of our interferometric data.
To determine the angular extent of the $K$-band emission, we analyze our
KI data in the context of a circularly symmetric uniform ring model 
\citep[e.g.,][]{EISNER+03}.  This simple model has been
shown \citep[e.g.,][]{EISNER+04} to provide a good description of more
complex models of inner disks including puffed-up inner rims 
\citep{DDN01,IN05}.  Furthermore, ring models have been used by
previous investigators \citep[e.g.,][]{MILLAN-GABET+07}, and our use
of this model will facilitate comparisons with previous results.

\begin{figure}
\plotone{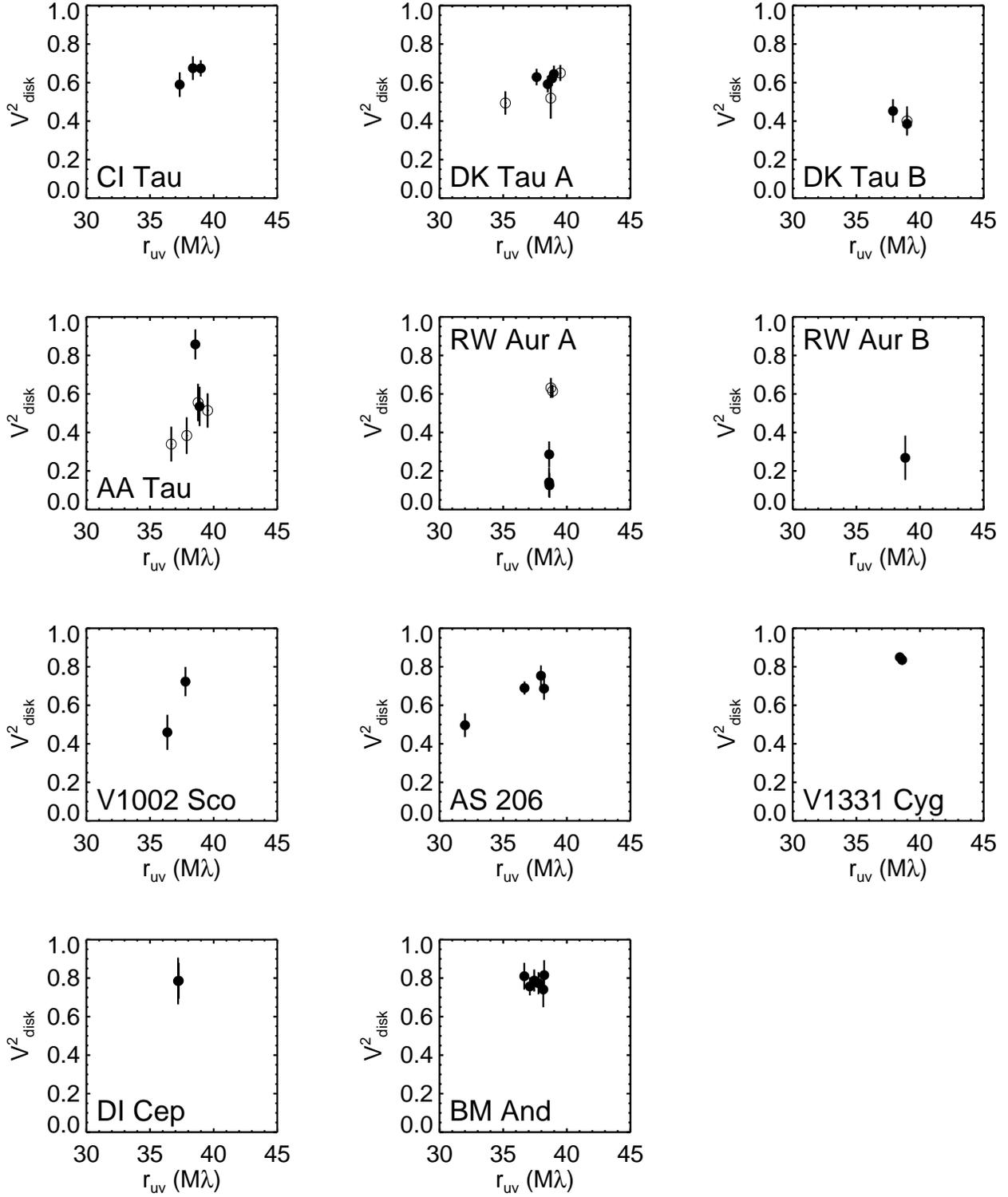}
\caption{Circumstellar component of the measured $V^2$ from KI 
($V^2_{\rm disk}$ as computed from Equation \ref{eq:v2disk}).
These squared visibilities represent the disk component only.  
Plotted error bars include uncertainties in measured $V^2$ and in
circumstellar-to-stellar flux ratios.  Different epochs are indicated
with different symbols: unfilled circles represent the second epoch
(where available).  The first epoch of KI data for RW Aur A are from
\citet{AKESON+05b}.
\label{fig:kidata}}
\end{figure}

We first remove the contribution to our measured $V^2$ from the
unresolved central star.  As described by \citet{EISNER+05}, decomposing
the observed $V^2$ into stellar and circumstellar components
requires a measurement of the circumstellar-to-stellar flux ratio;
the veilings determined in \S \ref{sec:rk} provide this ratio for our sample.
Contributions from the stellar and circumstellar components add linearly to 
the measured visibilities, and one can write,
\begin{equation}
V^2_{\rm disk} = \left\{\frac{\sqrt{V^2_{\rm meas}}(1+F_{\rm disk}/F_{\ast})-1}
{F_{\rm disk}/F_{\ast}}\right\}^2,
\label{eq:v2disk}
\end{equation}
where $F_{\rm disk}/F_{\ast}$ is the circumstellar-to-stellar flux ratio.
The $V^2_{\rm disk}$ values for our sample, obtained using the veiling
values measured closest in time to the measured $V^2$, are plotted in Figure 
\ref{fig:kidata}.

Uncertainties in our inferred circumstellar-to-stellar
flux ratios lead to uncertainties in $V^2_{\rm disk}$
in addition to those associated with $V^2_{\rm meas}$.  This error 
is smaller when the circumstellar-to-stellar flux ratio is large. Given 
inferred veilings and associated uncertainties for our sample, the resulting 
additional uncertainty is less than a few percent.  For the first epoch of 
data from RW Aur A, adopted from \citet{AKESON+05b}, 
the uncertainty is likely to be be larger (\S \ref{sec:rk}).

When decomposing observed $V^2$ values into
stellar and circumstellar parts, each epoch of KI observations was matched to 
the $K$-band veiling measurement closest in time.  Because our interferometric
and spectroscopic observations are not exactly simultaneous (generally, 
measurements are separated by a $\sim 1$ month), and objects may vary 
photometrically (at both optical and near-IR wavelengths) on this time-scale 
\citep[e.g.,][]{HERBST+94,SKRUTSKIE+96,EIROA+02},
the inferred
circumstellar-to-stellar flux ratios may not represent the true values
at the time of the KI observations.  Given this potential variability,
we also decomposed our multi-epoch $V^2_{\rm meas}$ into stellar and
circumstellar components by adopting the average value of the veiling over
all observed epochs.

The circumstellar components of the visibilities, $V^2_{\rm disk}$, were
fitted with uniform ring models using a $\chi^2$ minimization
to find the model providing the lowest residuals between predicted and
measured $V^2$.   
For DK Tau A,B, AA Tau, and RW Aur A, we fitted the ring model to each epoch 
of our KI observations.  
Best-fit radii for these face-on uniform ring models, using both the 
``nearest-in-time'' and ``average value'' circumstellar-to-stellar flux ratio
determinations,
are listed in Table \ref{tab:uds}.  Uncertainties listed in the table are
1$\sigma$ statistical uncertainties in the fits, obtained by computing
the surface where $\chi^2 = \chi_{\rm min}^2 + 1$.  Both approaches
to estimating the circumstellar-to-stellar flux ratios for multi-epoch
observations yield sizes consistent within 1$\sigma$ uncertainties.
In the analysis that follows, we will use the sizes derived with
the nearest-in-time veiling measurements.

Figure \ref{fig:kidata} shows that 
the $V^2$ for some objects (AA Tau, DK Tau A, RW Aur A, V1002 Sco, and AS 206)
do not decrease with increasing projected baseline length; these data
contrast with the predictions of circularly symmetric disk models. 
Photometrically determined stellar rotation periods combined with 
$v \sin i$ measurements suggest highly inclined geometries for some
of these sources
\citep[e.g.,][]{BOUVIER+95}, consistent 
with the non-monotonic behavior of $V^2$ versus baseline length in 
Figure \ref{fig:kidata}.  However,  because we are using a single-baseline 
interferometer (with a fixed separation and position angle), our data 
generally lack sufficient position angle coverage to place meaningful 
constraints on inclination (even accounting for Earth rotation, 
which helps to fill in the position angle coverage).

If we fit our data (assuming that multiple epochs of data
trace a non-variable source) with an inclined uniform disk model 
\citep[see e.g.,][]{EISNER+03}, we generally can not rule out an inclination
of $0^{\circ}$ (within 1-$\sigma$ uncertainties\footnote{For 3-parameter
(inner disk radius, position angle, and inclination) fits, 1$\sigma$
uncertainties are computed from projections of the surface with
$\chi^2=\chi_{\rm min}^2+3.53$.}).  Thus, we can not tell
whether the non-monotonic behavior seen in Figure \ref{fig:kidata}
is due to asymmetric source geometries or uncertainties in the measurements.
For one source, RW Aur A, our data are sufficient to constrain the inclination.
Assuming the source geometry has not varied over the two epochs of observation 
(see \S \ref{sec:var} for discussion of this assumption), the best-fit inclined
disk model has an inclination of $77^{+13}_{-15}$ degrees.

\subsection{Accretion Luminosities \label{sec:mdot}}
One of our aims in this analysis is to compare inferred inner disk properties
with stellar and accretion properties.  Accretion luminosity is
a particularly interesting property to correlate with our inner disk
measurements, since accretion can provide additional heating that modifies
the inner disk structure.  Here we use the Brackett gamma (Br$\gamma$) 
emission observed in
our NIRSPEC data as a diagnostic of infall \citep{MHC98,MCH01},
and thereby estimate accretion luminosities. 

\begin{figure}
\plotone{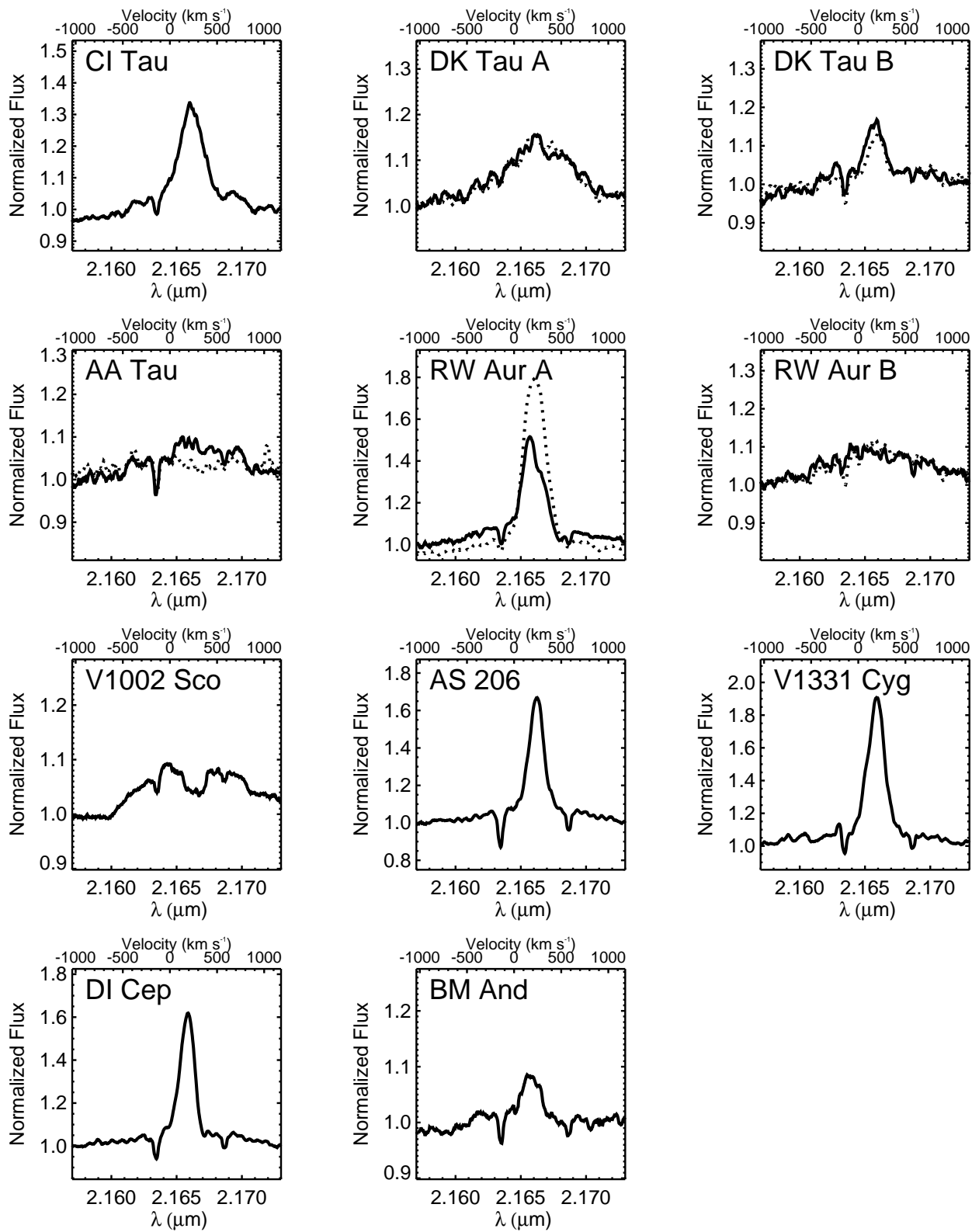}
\caption{Br$\gamma$ lines observed for our sample.  For objects 
observed over multiple epochs, the line observed during the first epoch is 
plotted with a solid line, and the line observed during the second
epoch is represented with a dotted line.
\label{fig:brg}}
\end{figure}

Br$\gamma$ emission lines for our observed sources are shown in Figure 
\ref{fig:brg}.  We measure the Br$\gamma$ equivalent widths (EW) of our targets
from these NIRSPEC observations by
1) determining the spectral baseline by fitting a straight line to a version of
the spectrum that is interpolated over the Br$\gamma$ feature;
2) removing this baseline from the (un-smoothed) target spectrum; 
and 3) integrating under the Br$\gamma$ line (over wavelengths from 2.157 to 
2.175 $\mu$m). We do not need to account for veiling here, since we are 
ultimately interested in the total line luminosity, which depends on the line 
EW as well as the continuum level (i.e., the correction to the EW
due to veiling would be offset by de-veiling the continuum when computing
the line flux).


Uncertainties in these EW measurements arise from noise in the spectra
and residual errors in calibration due to imperfect interpolation over the 
broad Br$\gamma$ absorption features in the telluric standards.
We assess these uncertainty sources by applying the same EW 
measurement procedure to non-accreting, weak-line T Tauri stars.  We find 
that the 1$\sigma$ uncertainty level for EW estimates is 0.1--0.2 
\AA \ for most sources.  For one object, V1002 Sco, the calibration did not 
work as well, and the error is 6 \AA. EWs and error bars for each source and 
epoch of observation are listed in Table \ref{tab:ews}.

We use the Br$\gamma$ EW measured from 
our NIRSPEC data and the $K_{\rm s}$-band 
magnitude determined from our photometry to estimate the Br$\gamma$ line flux.
For close binaries, where our near-IR photometry measures composite fluxes
of the system, we estimate fluxes of individual components using 
previously measured component flux ratios  \citep{SIMON+95,WG01}, which 
involves the implicit assumption that both component fluxes vary if the 
composite flux varies.  We de-redden our measured $K_{\rm s}$-band magnitudes 
using $A_V$ values from the literature \citep{WG01,EIROA+02,WALTER+94,HP92} and
the reddening law of \citet{ST91}, under the assumption that $A_V$ is not
time-variable.  We convert these de-reddened magnitudes
into units of flux density, $F_{\lambda}$.  We then compute the flux in the 
Br$\gamma$ line by multiplying $F_{\lambda}$  by the equivalent width:
$F_{\rm Br \gamma}=F_{\lambda} \times {\rm EW}$.  
The Br$\gamma$ luminosity is 
then $L_{\rm Br \gamma}=4 \pi d^2 F_{\rm Br \gamma}$, where $d$ is the
distance.  The Br$\gamma$ luminosity is related to the accretion luminosity
by an empirical relation from \citep{MHC98,MCH01}:
$\log(L_{\rm acc}/L_{\odot}) = (1.26 \pm 0.19)\log(L_{\rm Br \gamma}/L_{\odot})
+(4.43 \pm 0.79)$.  

Inferred values of $L_{\rm Br \gamma}$ and $L_{\rm acc}$ are included
in Table \ref{tab:ews}.  Our estimated $L_{\rm acc}$ values are compatible
with previous measurements from the literature 
\citep[e.g.,][]{VBJ93,HEG95,GULLBRING+98,CG98,WG01}.  Previous measurements
typically span an order of magnitude, and our inferred $L_{\rm acc}$ are within
an order of magnitude of all previously determined values.  Moreover, the
values for each star relative to others in our sample roughly match those 
of surveys in the literature; in surveys where several of our targets 
were observed, the lowest accretion rate objects are also the lowest accretors
in that subset of our sample.

\section{Discussion}

\subsection{Size-Luminosity Diagram}
If dusty disks extended inward to their central stars, their near-IR emission
would be very compact, so much so that it would appear unresolved even with
the mas-level angular resolution of near-IR interferometers.  The fact that
disks around young stars {\it are} resolved with near-IR interferometric data
\citep[e.g.,][]{MILLAN+99}
indicates that these disks are truncated at stellocentric radii substantially 
larger than the stellar radius.  Measured temperatures at the disk
truncation radii are typically $\sim 1500$--2000 K 
\citep[e.g.,][]{EISNER+04,EISNER+05},
compatible with the sublimation temperature of silicate dust 
\citep[e.g.,][]{POLLACK+94}.  Determining how the inner disk truncation radius
depends on source luminosity can constrain how the inner edge is heated and,
in turn, the radial and vertical structure of the inner disk
\citep[e.g.,][]{MM02,EISNER+04}.

\begin{figure}
\plotone{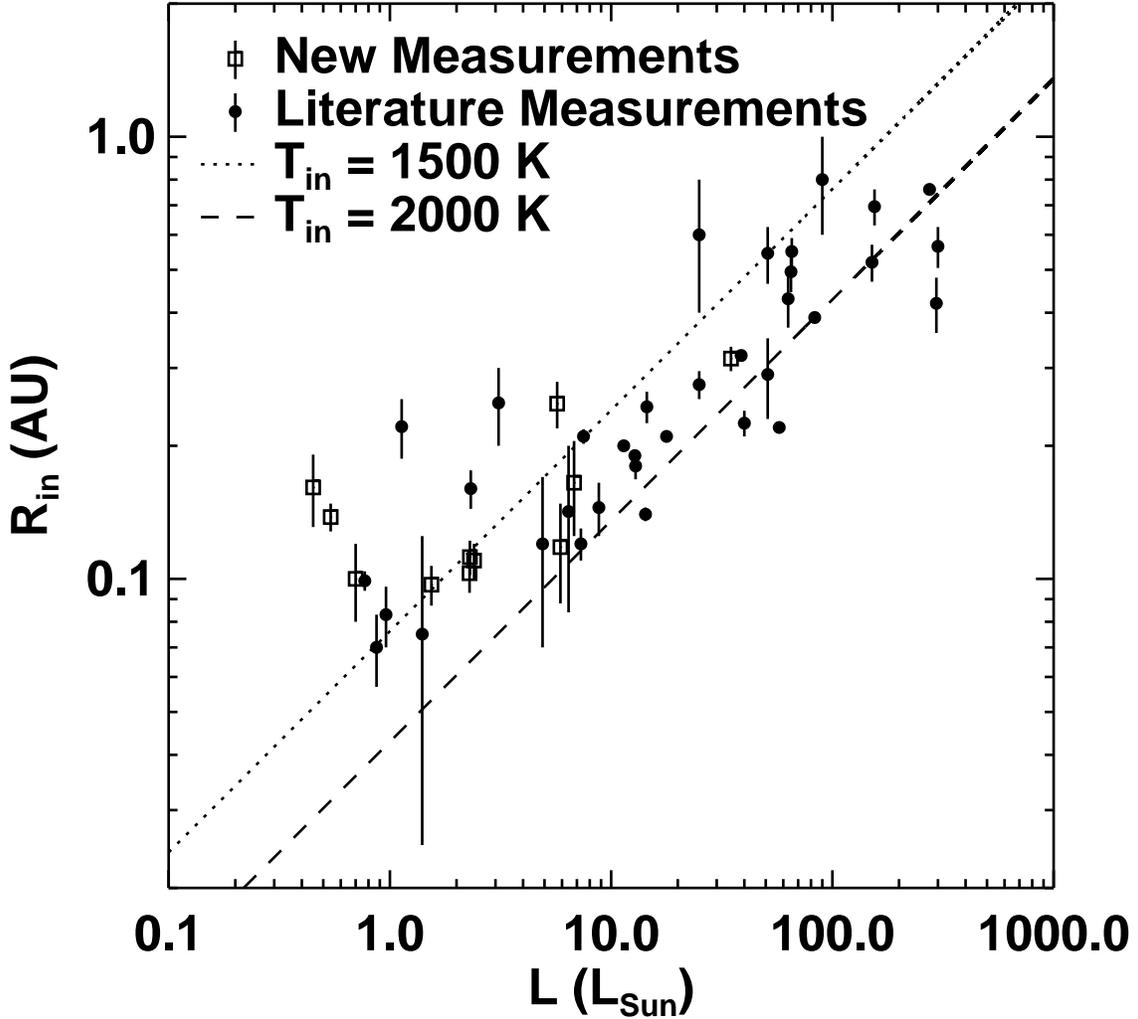}
\caption{Inner ring radii determined from near-IR interferometry, as
a function of source luminosity (stellar+accreion luminosities, where 
available).  Inner ring sizes
from the literature ({\it filled circles}) and for the sources in our sample
({\it open squares}) are plotted.
We indicate the inner ring radii
expected for a puffed-up inner disk model \citep{DDN01}
with $T_{\rm in}=1500$ K
({\it dotted line}) and $T_{\rm in}=2000$ K ({\it dashed line}).
While most objects with luminosities $\ga 10$ L$_{\odot}$ agree with
the predictions of these models \citep[with the excpetion of several high-mass
stars whose disks appear consistent with geometrically thin disk models;][]
{EISNER+04}, at lower luminosities, there are
numerous sources with inner ring radii larger than predicted values.
\label{fig:sizes}}
\end{figure}

Figure \ref{fig:sizes} shows the inner ring radii determined for a sample
of T Tauri and Herbig Ae/Be objects (including 
previous measurements from the literature) as a function of source
luminosity.  The source luminosity is the sum of the stellar luminosity and
the accretion luminosity; in cases where no measurement of accretion luminosity
is available (mostly high-mass stars), only the stellar luminosity is used.
Inner ring sizes from the literature are drawn from 
\citet{MST01,EISNER+04,EISNER+05,AKESON+00,AKESON+05b,AKESON+05,MONNIER+05};
we used luminosities for these sources quoted in these references.
For the objects in our current sample, we took stellar
luminosities from \citet{KH95,WG01,WALTER+94,HP92}, and accretion luminosities
from Table \ref{tab:ews}.  For objects where we measured the inner disk size
over multiple epochs, the average size is plotted in Figure \ref{fig:sizes}.

We compare the size-luminosity diagram in Figure \ref{fig:sizes} with the
predictions of a simple physical model.  Specifically, we assume that the
inner disk is directly irradiated by the central star, and that the
inner edge therefore puffs up \citep{DDN01}.  For this model, the location
of the inner disk edge is,
\begin{equation}
R_{\rm in} = \sqrt{(1+f) \left(\frac{L_{\ast}+L_{\rm acc}}{4 \pi \sigma 
T_{\rm in}^4}\right)}.
\label{eq:ddn}
\end{equation}
Here, $f$ is the ratio of the inner edge height to its radius, 
and is estimated to be between 0.1-0.2 for T Tauri and 
Herbig Ae/Be stars \citep{DDN01}.
$T_{\rm in}$ is the temperature at the inner disk edge.  Figure \ref{fig:sizes}
shows the inner disk radii predicted by this model for $T_{\rm in}=1500$ K
and $T_{\rm in}=2000$ K; these temperatures bracket the expected sublimation
temperatures for silicate dust in a protoplanetary disk 
\citep[e.g.,][]{POLLACK+94}.

Most objects with luminosities 
$\ga 10$ L$_{\odot}$ agree with the predictions of these models; this fact has
been noted previously \citep{MM02,EISNER+04,MILLAN-GABET+07}.
There are, however, several exceptional high luminosity sources for which 
geometrically thin disk models may be more suitable 
\citep[e.g.,][]{EISNER+04,MM02,VJ07}.
Moreover, Figure \ref{fig:sizes} shows that at the lowest stellar masses, 
this model may break down.  The sizes of several low-mass sources are larger 
than predicted for these puffed-up inner disk models.  

We put forward several potential explanations for these deviations between 
puffed-up disk model predictions and data for low-mass T Tauri stars.  
First, the sublimation temperature may be reached further from the star.
For example, viscous energy dissipation in accreting material may provide
additional heating, pushing the sublimation radius outward.  The
sublimation temperature may also be lower, and hence located at larger
stellocentric radii, in these low-mass objects because of 
systematically lower ambient gas densities
\citep[see Table 3 of][]{POLLACK+94}.  Small dust grains, which are hotter 
than larger grains at a given radius, may also lead to larger dust
sublimation radii.  Finally, these disks may be truncated
by a mechanism other than dust sublimation, such as photoevaporative mass loss
or magnetospheric truncation.  In the following paragraphs,
we argue that of the possible explanations considered, the most likely 
is a mechanism other than dust sublimation, namely, magnetospheric
truncation.

\begin{figure}
\plotone{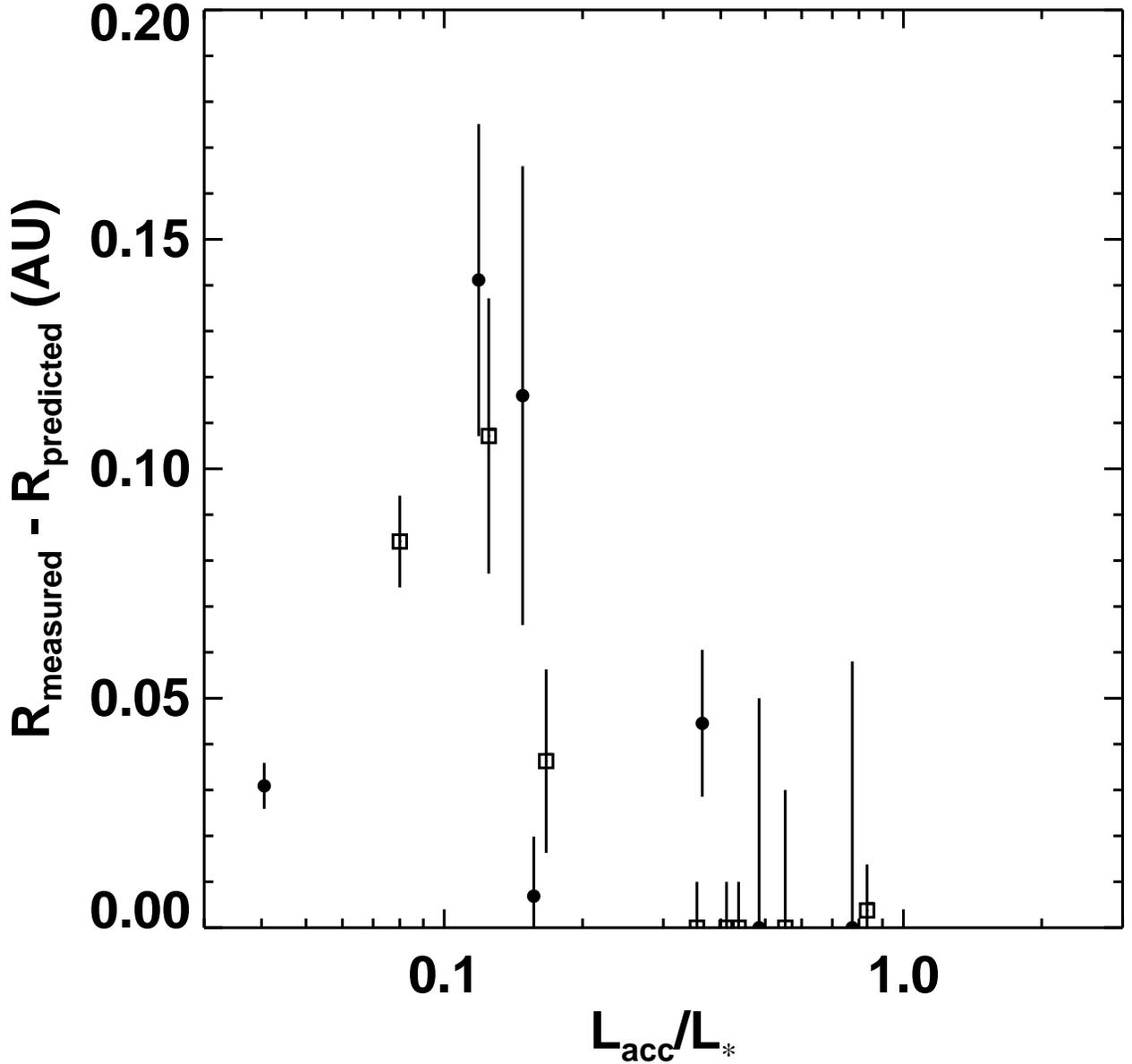}
\caption{
Discrepancy between the measured size and the
model prediction (in the range of inner disk temepratures between 1500 and
2000 K) is plotted against the ratio of accretion to stellar luminosity.
The discrepancy is the difference between the measured size and
the region bounded by the two curves in Figure \ref{fig:sizes}. 
As in Figure \ref{fig:sizes}, disrepancies for sources
from the literature are indicated by filled circles and, for the sources in our
sample, by open squares.
We have restricted this plot to T Tauri stars
($L_{\ast} < 5$ L$_{\odot}$), since most higher luminosity
objects do not have measured $L_{\rm acc}$.
While some low accretion luminosity sources do agree with the model
predictions, a trend is nevertheless evident whereby objects with smaller 
$L_{\rm acc}/L_{\ast}$ ratios are more discrepant from model predictions.
The objects exhibiting the largest discrepancies are DK Tau B and
RW Aur B from our sample, GM Aur from \citet{AKESON+05b}, and AS 207 A from
\citet{EISNER+05}.
\label{fig:laccs}}
\end{figure}

These various possible explanations make different predictions about the
accretion rates of sources with exceptionally large inner disk radii.
We therefore investigate how the discrepancy between measured sizes and 
predictions of irradiated disk models depends
on a proxy of source accretion rate\footnote{We use $L_{\rm acc}$ rather
than $\dot{M}$, since $L_{\rm acc}$ is closer to the observed quantities.
Conversion to $\dot{M}$ requires additional assumptions about stellar
mass and radius, and the inner disk radius \citep[e.g.,][]{GULLBRING+98}.}.  
Figure \ref{fig:laccs} plots the 
difference between
measured and predicted size as a function of $L_{\rm acc}/L_{\ast}$.
We find that objects whose measured sizes are larger than predicted by models
actually have lower (fractional) accretion luminosities than other sources. 

\subsubsection{Accretional Heating \label{sec:acc}}
If accretional heating causes larger inner disk truncation radii,
then the low-mass stars in Figure \ref{fig:sizes} would have systematically 
higher mass accretion rates than other sources.   However, Figure
\ref{fig:laccs} shows that objects with large inner radii tend to have
lower accretion rates.  Thus, 
accretional heating seems an unlikely explanation for the large inner
disk sizes.

\subsubsection{Lower Dust Sublimation Temperatures \label{sec:rsub}}
Dust sublimates more easily when ambient pressures are lower.  Lower
mass accretion rates lead to lower gas densities and pressures, and
can therefore yield lower sublimation temperatures. The fact that larger disk 
truncation radii tend to occur around sources with lower accretion 
luminosities (Figure \ref{fig:laccs}) is compatible with the hypothesis
that these disks are truncated at lower dust sublimation temperatures.

However, the dust sublimation temperature
would need to be $\la 1000$ K to explain many of the large measured disk
sizes. A silicate dust sublimation temperature of 1000 K requires 8 orders of 
magnitude lower gas density than a sublimation temperature of 1500 K
\citep{POLLACK+94}, which corresponds to $\sim 5$ orders of magnitude
lower mass accretion rates (assuming an $\alpha$-disk to convert gas density
into mass accretion rate).  Such low accretion rates, and large dispersion in 
accretion rate, are not measured for our sample.  It is therefore
unlikely that lower sublimation temperature can explain the large
truncation radii of disks around low-mass T Tauri stars.

\subsubsection{Small Dust Grains \label{sec:agrain}}
Smaller dust grains absorb and emit radiation less efficiently than larger 
grains, and this loss of efficiency is more severe at longer wavelengths
\citep[e.g.,][]{BH83}.  As a 
result, small grains will achieve higher temperatures than larger grains
at a given stellocentric radius.  Thus, if a disk is composed of small grains, 
the dust sublimation radius can be larger than implied by Equation 
\ref{eq:ddn}, which assumed gray, and hence large-grained, dust 
\citep{MM02,ITN06,VINKOVIC+06,THM07}.
Using the results of \citet{VINKOVIC+06}, we find that the large inner
disk sizes in our sample can be reproduced by puffed up inner disk models
composed of grains sized 0.1--0.5 $\mu$m.
While small-dust-grain models can reproduce these larger observed sizes,
it seems unlikely that those objects that would  require small dust grains to 
explain the observations are precisely those with the lowest mass accretion 
rates.  

\subsubsection{Photo-evaporation \label{sec:photo}}
We now consider alternatives to dust sublimation, where the
physical process responsible for truncating the disk might be more
efficient at lower mass accretion rates.  One such process is
photo-evaporation, where the mass-loss driven by stellar UV radiation
can overcome accretion at sufficiently low $\dot{M}$ 
\citep[e.g.,][]{CGS01,ACP06}.
The mass flow can ``switch'' from accretion to outflow when the accretion
rate drops below $\sim 10^{-10}$ M$_{\odot}$ yr$^{-1}$.  Since
the objects in our sample have accretion rates $\ga 10^{-9}$ M$_{\odot}$
yr$^{-1}$ (based on the accretion luminosities in Table \ref{tab:ews}),
photo-evaporation is unlikely to dominate accretion.  Moreover, we do
not see the signature of a discrete transition in Figure \ref{fig:laccs}
at a certain accretion luminosity.  Thus, photo-evaporation is unlikely
to be the physical mechanism behind the truncated disks observed
for the low-mass T Tauri stars in our sample.

\subsubsection{Magnetospheric Truncation \label{sec:rmag}}
Large inner disk radii may also be due to magnetospheric truncation.
\mbox{T Tauri} 
stars accrete material via magnetospheric accretion; the stellar magnetic
field threads the disk and essentially creates a barrier to midplane
accretion at the point of balance between magnetic pressure and 
inward pressure from accretion \citep[e.g.,][]{SHU+94}.  
The magnetospheric radius is given by \citep{KONIGL91}:
\begin{equation}
\frac{R_{\rm mag}}{R_{\ast}} = 2.27 \left[\frac{(B_0/{\rm 1 \: kG})^4 
(R_{\ast}/{\rm R_{\odot}})^5}{(M_{\ast}/{\rm M_{\odot}}) 
(\dot{M}/10^{-7} {\rm M_{\odot} \: yr}^{-1})^2}\right]^{1/7},
\label{eq:rmag}
\end{equation}
where $B_0$ is the dipole magnetic field strength.
For many classical T Tauri stars, high disk accretion rates ($\ga 10^{-7}$
M$_{\odot}$ yr$^{-1}$) confine this magnetospheric 
radius to well within the dust sublimation radius \citep[e.g.,][]{EISNER+05},
and thus the observed inner edge of the dust disk occurs where
temperatures reach the sublimation temperature (i.e., $T_{\rm in}=1500$--2000 
K).  

However, lower accretion rates allow the magnetospheric radius to
extend to larger stellocentric radii ($R_{\rm mag} \propto \dot{M}^{-2/7}$), 
while the lower associated accretion luminosity leads to a smaller 
stellocentric radius of dust sublimation.  Low accretion rates can therefore
cause magnetospheric truncation outside of the
dust sublimation radius \citep[e.g.,][]{ECH06}.  If we take 
$B_0 = 1$ kG, $R_{\ast}=2$ R$_{\odot}$, and $M_{\ast}=0.5$ M$_{\odot}$
as typical values for low-mass T Tauri stars \citep[e.g.,][]{JOHNS-KRULL07},
then mass accretion rates between $10^{-9}$ and $10^{-8}$ M$_{\odot}$ yr$^{-1}$
(compatible with the lowest accretion luminosities in Table \ref{tab:ews})
yield magnetospheric radii between 0.07 and 0.15 AU, comparable to the
measured inner disk sizes in Figure \ref{fig:sizes}.  We thus favor
magnetospheric truncation as the explanation for the large inner disk sizes of
low-mass T Tauri stars.

\subsection{Inner Disk Variability \label{sec:var}}
Many young stars, including those in our sample, are known to be 
photometrically variable at optical through near-infrared wavelengths,
with variability timescales ranging from days to months
\citep[e.g.,][]{HERBST+94,SKRUTSKIE+96,CHS01,EIROA+02,BOUVIER+07}.  While the 
underlying cause of the variability is usually attributed to star-spots, which 
modulate the emission on the stellar rotation period, in some cases inner disk
variability may also play a role \citep[e.g.,][]{EIROA+02,BOUVIER+07}.  

Time-varying increases in disk accretion rates, which might occur on 
inner disk dynamical timescales of days to weeks, could lead to increased disk 
heating and higher near-IR emission.  Since a hotter disk 
could sublimate dust at larger stellocentric radii than a cooler disk, the 
inner edge of the dust disk may move in and out as the disk heating varies. 
Conversely, if disks are magnetospherically truncated, then increases
in accretion rate would lead to smaller disk sizes (Equation \ref{eq:rmag}).
Variability in the stellar magnetic field strength can also cause variability 
in mass accretion rates and 
changes in the inner disk radius over timescales comparable
to stellar dynamo oscillation period, typically $\sim 10$ yr 
\citep[e.g.,][]{ARMITAGE95}.

We obtained multiple epochs of spectroscopic and photometric data for 
DK Tau A and B, AA Tau,
and RW Aur A and B, and multiple epochs of interferometric data for all of 
these except RW Aur B.  
As indicated in Tables \ref{tab:phot}--\ref{tab:ews}, several sources
have variable infrared fluxes, veilings, and inferred accretion
luminosities.  Moreover, one object (RW Aur A) may have a variable inner disk
size (Figure \ref{fig:kidata}; Table \ref{tab:uds}).  

In the scenario where observed variability is due to changes in the disk
heating (e.g., because of time-variable disk accretion rates), we would
expect the following for higher disk temperatures: 1) $K$-band flux will
increase; 2) veiling will increase, 
since the circumstellar flux is increasing relative to the stellar flux;
3) accretion luminosity may increase (if this is the source of the disk
heating); 4) the measured inner disk size should increase, since the
dust sublimation radius would move out.  

None of our targets clearly show such signatures.  For DK Tau A, 
the $K$-band veiling, near-IR fluxes, and inner disk size 
remain constant (within uncertainties) over the span of our multi-epoch
observations. For AA Tau, the veiling appears to decrease 
from one epoch to the next while the Br$\gamma$ flux also decreases.  However, 
the inner disk size remains constant (or marginally increases; 
Table \ref{tab:uds})) 
and the $K_{\rm s}$-band flux appears to 
increase in the latter epoch, contrary to expectations for a dimmer disk. For 
\mbox{RW Aur A}, the Br$\gamma$ luminosity appears to increase as the veiling
decreases, which is not consistent with expectations for our scenario
of inner disk variability above.  DK Tau B and RW Aur B show no variability
in any of our data.

If the emission from our targets is circularly symmetric, then our modeling
indicates potential variability (above the 1$\sigma$ level) in the inner disk 
size with time for \mbox{RW Aur A}.  
The inner disk radius appears to increase as 
the $K$-band veiling increases, which is expected since a larger inner rim
would produce more near-IR emission (for a fixed temperature).  
However, Figure \ref{fig:raw} shows only marginal variability 
in the measured $V^2$ (in contrast to the $V^2_{\rm disk}$ plotted in 
Figure \ref{fig:kidata}) from one epoch to the next, and 
the apparent inner disk 
variability rests entirely on the inferred circumstellar-to-stellar flux 
ratios.  The first epoch of KI data for \mbox{RW Aur A} does not have
a reliable estimate of this ratio; the estimate of 0.6 from \citet{AKESON+05b}
is a lower limit (\S \ref{sec:rk}), and a ratio as high as 2.5 would eliminate 
the apparent variability.
Furthermore, a highly inclined disk can fit the data for 
\mbox{RW Aur A} well without any variability 
in the inner disk size, even if the inferred circumstellar-to-stellar 
flux ratios are taken at face value (\S \ref{sec:rin}).
Thus we can not claim to see inner disk variability based on these data.

The interpretation of our multi-epoch observations is complicated by the fact
that different data constituting each epoch were not obtained simultaneously.
Data for a given epoch were obtained within a few week time-window.  Because 
the dynamical timescales for inner disk variability are 
$\tau_{\rm dyn} \sim \sqrt{R^3/GM} \la 10$ days for the inner radii 
inferred for our sample, there may be substantial source variability between 
the different measurements within a single epoch.  In future work, the 
observations constituting a single epoch should ideally be obtained within 
$\sim 1$ day of one another.  In addition, a larger number of epochs would
facilitate investigation of inner disk variability.  

\section{Conclusions}
Using near-IR interferometric, spectroscopic, and photometric data, we
measured the inner disk radii of 11 low-mass T Tauri stars.  In addition,
we measured the near-IR veilings and accretion 
luminosities of these objects.   Our data substantiate previous claims
that the inner disk radii of T Tauri stars are generally consistent with the 
predictions of disk models with vertically extended inner rims at 
stellocentric radii where disk temperatures reach dust sublimation 
temperatures, between 1500 and 2000 K.  However, there are a handful of 
sources, in particular objects with stellar luminosities $\la 1$ L$_{\odot}$,
for which the measured radii are substantially larger than predicted by
such models. 

Discrepancies between models where disks are truncated at the
dust sublimation radius and measured inner disk radii are larger for sources
with low ratios of accretion to stellar luminosity.  Thus, the physical 
mechanism by which the inner disk is truncated appears to depend on 
the relative
importance of stellar irradiation and accretion.  To explain this observation,
we suggest that at low accretion rates, as the pressure of accreting
material drops, the point of balance between inward accretion pressure
and outward stellar magnetic pressure moves to larger stellocentric
radii. At sufficiently low accretion rates, magnetospheric radii
can become larger than the dust sublimation radii, yielding the large
observed sizes for low-accretion-rate objects in our sample.

For several of our sample objects, we obtained multiple epochs of 
interferometric, spectroscopic, and photometric data, and for each epoch
the various datasets were obtained within a few weeks of one another.
We used these multi-epoch observations to search for inner disk variability.
While inferred veilings, accretion luminosities, and even inner disk truncation
radii, appear to vary for some sources, none of the objects show a
variability signature consistent with changes in the inner disk.
Future monitoring observations with more epochs, and with different data 
comprising each epoch obtained within a time-window short compared to inner
disk dynamical timescales, will enable a more
rigorous probe of inner disk variability.

\bigskip
The near-IR interferometry data presented in this paper were obtained
with the Keck Interferometer (KI), which is supported by NASA.  We wish
to thank the entire KI team for making these observations possible.
KI and NIRSPEC observations were carried out at the W.M. Keck Observatory, 
which is operated as a scientific partnership among California Institute of 
Technology, the University of California, and NASA.  The Observatory was made 
possible by the generous financial support of the W.M. Keck Foundation.
The authors wish to recognize and acknowledge the cultural role and reverence 
that the summit of Mauna Kea has always had within the indigenous Hawaiian 
community. We are most fortunate to have the opportunity to conduct 
observations from this mountain.  
This work has used software from the Michelson Science Center at
the California Institute of Technology. J.A.E. acknowledges 
support from a Miller Research Fellowship, and thanks G. Basri for useful
discussions and input into various stages of this work.  J.A.E. is also
grateful to F. Ciesla for interesting discussion about dust sublimation 
temperatures, and to E. Lopez for his interest in this project.








\bibliographystyle{apj}

\begin{thebibliography}{76}

\bibitem[{{Adams} {et~al.}(1988){Adams}, {Lada}, \& {Shu}}]{ALS88}
{Adams}, F.~C., {Lada}, C.~J., \& {Shu}, F.~H. 1988, \apj, 326, 865

\bibitem[{{Akeson} {et~al.}(2005{\natexlab{a}}){Akeson}, {Boden}, {Monnier},
  {Millan-Gabet}, {Beichman}, {Beletic}, {Calvet}, {Hartmann}, {Hillenbrand},
  {Koresko}, {Sargent}, \& {Tannirkulam}}]{AKESON+05b}
{Akeson}, R.~L., {Boden}, A.~F., {Monnier}, J.~D., {Millan-Gabet}, R.,
  {Beichman}, C., {Beletic}, J., {Calvet}, N., {Hartmann}, L., {Hillenbrand},
  L., {Koresko}, C., {Sargent}, A., \& {Tannirkulam}, A. 2005{{a}},
  \apj, 635, 1173

\bibitem[{{Akeson} {et~al.}(2000){Akeson}, {Ciardi}, {van Belle},
  {Creech-Eakman}, \& {Lada}}]{AKESON+00}
{Akeson}, R.~L., {Ciardi}, D.~R., {van Belle}, G.~T., {Creech-Eakman}, M.~J.,
  \& {Lada}, E.~A. 2000, \apj, 543, 313

\bibitem[{{Akeson} {et~al.}(2005{b}){Akeson}, {Walker}, {Wood},
  {Eisner}, {Scire}, {Penprase}, {Ciardi}, {van Belle}, {Whitney}, \&
  {Bjorkman}}]{AKESON+05}
{Akeson}, R.~L., {Walker}, C.~H., {Wood}, K., {Eisner}, J.~A., {Scire}, E.,
  {Penprase}, B., {Ciardi}, D.~R., {van Belle}, G.~T., {Whitney}, B., \&
  {Bjorkman}, J.~E. 2005{\natexlab{b}}, \apj, 622, 440

\bibitem[{{Alexander} {et~al.}(2006){Alexander}, {Clarke}, \&
  {Pringle}}]{ACP06}
{Alexander}, R.~D., {Clarke}, C.~J., \& {Pringle}, J.~E. 2006, \mnras, 369, 229

\bibitem[{{Armitage}(1995)}]{ARMITAGE95}
{Armitage}, P.~J. 1995, \mnras, 274, 1242

\bibitem[{{Aveni} \& {Hunter}(1969)}]{AH69}
{Aveni}, A.~F. \& {Hunter}, J.~H. 1969, \aj, 74, 1021

\bibitem[{{Basri} \& {Batalha}(1990)}]{BB90}
{Basri}, G. \& {Batalha}, C. 1990, \apj, 363, 654

\bibitem[{{Basri} \& {Bertout}(1989)}]{BB89}
{Basri}, G. \& {Bertout}, C. 1989, \apj, 341, 340

\bibitem[{{Beckwith} {et~al.}(1990){Beckwith}, {Sargent}, {Chini}, \&
  {Guesten}}]{BECKWITH+90}
{Beckwith}, S.~V.~W., {Sargent}, A.~I., {Chini}, R.~S., \& {Guesten}, R. 1990,
  \aj, 99, 924

\bibitem[{{Bertout} {et~al.}(1988){Bertout}, {Basri}, \& {Bouvier}}]{BBB88}
{Bertout}, C., {Basri}, G., \& {Bouvier}, J. 1988, \apj, 330, 350

\bibitem[{{Bloom} {et~al.}(2006){Bloom}, {Starr}, {Blake}, {Skrutskie}, \&
  {Falco}}]{BLOOM+06}
{Bloom}, J.~S., {Starr}, D.~L., {Blake}, C.~H., {Skrutskie}, M.~F., \& {Falco},
  E.~E. 2006, in Astronomical Data Analysis Software and Systems XV, ed.
  C.~{Gabriel}, C.~{Arviset}, D.~{Ponz}, \& S.~{Enrique}, Vol. 351, 751

\bibitem[{{Boden} {et~al.}(1998){Boden}, {Colavita}, {van Belle}, \&
  {Shao}}]{BODEN+98}
{Boden}, A.~F., {Colavita}, M.~M., {van Belle}, G.~T., \& {Shao}, M. 1998, in
  Proc. SPIE Vol. 3350, p. 872-880, Astronomical Interferometry, Robert D.
  Reasenberg; Ed., 872--880

\bibitem[Bohren \& Huffman(1983)]{BH83} Bohren, C.~F., \& 
Huffman, D.~R.\ 1983, New York: Wiley  

\bibitem[{{Bouvier} {et~al.}(2007){Bouvier}, {Alencar}, {Boutelier},
  {Dougados}, {Balog}, {Grankin}, {Hodgkin}, {Ibrahimov}, {Kun}, {Magakian}, \&
  {Pinte}}]{BOUVIER+07}
{Bouvier}, J., {Alencar}, S.~H.~P., {Boutelier}, T., {Dougados}, C., {Balog},
  Z., {Grankin}, K., {Hodgkin}, S.~T., {Ibrahimov}, M.~A., {Kun}, M.,
  {Magakian}, T.~Y., \& {Pinte}, C. 2007, \aap, 463, 1017

\bibitem[{{Bouvier} {et~al.}(1995){Bouvier}, {Covino}, {Kovo}, {Martin},
  {Matthews}, {Terranegra}, \& {Beck}}]{BOUVIER+95}
{Bouvier}, J., {Covino}, E., {Kovo}, O., {Martin}, E.~L., {Matthews}, J.~M.,
  {Terranegra}, L., \& {Beck}, S.~C. 1995, \aap, 299, 89

\bibitem[{{Calvet} \& {Gullbring}(1998)}]{CG98}
{Calvet}, N. \& {Gullbring}, E. 1998, \apj, 509, 802

\bibitem[{{Carpenter} {et~al.}(2001){Carpenter}, {Hillenbrand}, \&
  {Skrutskie}}]{CHS01}
{Carpenter}, J.~M., {Hillenbrand}, L.~A., \& {Skrutskie}, M.~F. 2001, \aj, 121,
  3160

\bibitem[{{Chavarria-K.}(1981)}]{CK81}
{Chavarria-K.}, C. 1981, \aap, 101, 105

\bibitem[{{Clarke} {et~al.}(2001){Clarke}, {Gendrin}, \& {Sotomayor}}]{CGS01}
{Clarke}, C.~J., {Gendrin}, A., \& {Sotomayor}, M. 2001, \mnras, 328, 485

\bibitem[{{Colavita} {et~al.}(2003){Colavita}, {Akeson}, {Wizinowich}, {Shao},
  {Acton}, {Beletic}, {Bell}, {Berlin}, {Boden}, {Booth}, {Boutell}, {Chaffee},
  {Chan}, {Chock}, {Cohen}, {Crawford}, {Creech-Eakman}, {Eychaner},
  {Felizardo}, {Gathright}, {Hardy}, {Henderson}, {Herstein}, {Hess},
  {Hovland}, {Hrynevych}, {Johnson}, {Kelley}, {Kendrick}, {Koresko}, {Kurpis},
  {Le Mignant}, {Lewis}, {Ligon}, {Lupton}, {McBride}, {Mennesson},
  {Millan-Gabet}, {Monnier}, {Moore}, {Nance}, {Neyman}, {Niessner}, {Palmer},
  {Reder}, {Rudeen}, {Saloga}, {Sargent}, {Serabyn}, {Smythe}, {Stomski},
  {Summers}, {Swain}, {Swanson}, {Thompson}, {Tsubota}, {Tumminello}, {van
  Belle}, {Vasisht}, {Vause}, {Walker}, {Wallace}, \& {Wehmeier}}]{COLAVITA+03}
{Colavita}, M.~M., et al. 2003, \apjl, 592, L83

\bibitem[{{Colavita}(1999)}]{COLAVITA99}
{Colavita}, M.~M. 1999, \pasp, 111, 111

\bibitem[{{Colavita} \& {Wizinowich}(2003)}]{CW03}
{Colavita}, M.~M. \& {Wizinowich}, P.~L. 2003, in Interferometry for Optical
  Astronomy II. Edited by Wesley A. Traub. Proceedings of the SPIE, Volume
  4838, pp. 79-88 (2003)., 79--88

\bibitem[{{Dullemond} {et~al.}(2001){Dullemond}, {Dominik}, \& {Natta}}]{DDN01}
{Dullemond}, C.~P., {Dominik}, C., \& {Natta}, A. 2001, \apj, 560, 957

\bibitem[{{Dutrey} {et~al.}(1996){Dutrey}, {Guilloteau}, {Duvert}, {Prato},
  {Simon}, {Schuster}, \& {Menard}}]{DUTREY+96}
{Dutrey}, A., {Guilloteau}, S., {Duvert}, G., {Prato}, L., {Simon}, M.,
  {Schuster}, K., \& {Menard}, F. 1996, \aap, 309, 493

\bibitem[{{Edwards} {et~al.}(1994){Edwards}, {Hartigan}, {Ghandour}, \&
  {Andrulis}}]{EDWARDS+94}
{Edwards}, S., {Hartigan}, P., {Ghandour}, L., \& {Andrulis}, C. 1994, \aj,
  108, 1056

\bibitem[{{Eiroa} {et~al.}(2002){Eiroa}, {Oudmaijer}, {Davies}, {de Winter},
  {Garz{\' o}n}, {Palacios}, {Alberdi}, {Ferlet}, {Grady}, {Cameron}, {Deeg},
  {Harris}, {Horne}, {Mer{\'{\i}}n}, {Miranda}, {Montesinos}, {Mora}, {Penny},
  {Quirrenbach}, {Rauer}, {Schneider}, {Solano}, {Tsapras}, \&
  {Wesselius}}]{EIROA+02}
{Eiroa}, C., et al. 2002, \aap, 384, 1038

\bibitem[{{Eisner} {et~al.}(2006){Eisner}, {Chiang}, \& {Hillenbrand}}]{ECH06}
{Eisner}, J.~A., {Chiang}, E.~I., \& {Hillenbrand}, L.~A. 2006, \apjl, 637,
  L133

\bibitem[{{Eisner} {et~al.}(2007){Eisner}, {Chiang}, {Lane}, \&
  {Akeson}}]{EISNER+07a}
{Eisner}, J.~A., {Chiang}, E.~I., {Lane}, B.~F., \& {Akeson}, R.~L. 2007, \apj,
  657, 347

\bibitem[{{Eisner} {et~al.}(2005){Eisner}, {Hillenbrand}, {White}, {Akeson}, \&
  {Sargent}}]{EISNER+05}
{Eisner}, J.~A., {Hillenbrand}, L.~A., {White}, R.~J., {Akeson}, R.~L., \&
  {Sargent}, A.~I. 2005, \apj, 623, 952

\bibitem[{{Eisner} {et~al.}(2003){Eisner}, {Lane}, {Akeson}, {Hillenbrand}, \&
  {Sargent}}]{EISNER+03}
{Eisner}, J.~A., {Lane}, B.~F., {Akeson}, R.~L., {Hillenbrand}, L., \&
  {Sargent}, A. 2003, \apj, 588, 360

\bibitem[{{Eisner} {et~al.}(2004){Eisner}, {Lane}, {Hillenbrand}, {Akeson}, \&
  {Sargent}}]{EISNER+04}
{Eisner}, J.~A., {Lane}, B.~F., {Hillenbrand}, L., {Akeson}, R., \& {Sargent},
  A. 2004, \apj, 613, 1049

\bibitem[{{Gray}(1992)}]{GRAY92}
{Gray}, D.~F. 1992, {The Observation and Analysis of Stellar Photospheres} 
  (Cambridge University Press, UK)

\bibitem[{{Gullbring} {et~al.}(1998){Gullbring}, {Hartmann}, {Briceno}, \&
  {Calvet}}]{GULLBRING+98}
{Gullbring}, E., {Hartmann}, L., {Briceno}, C., \& {Calvet}, N. 1998, \apj,
  492, 323

\bibitem[{{Hamann} \& {Persson}(1992)}]{HP92}
{Hamann}, F. \& {Persson}, S.~E. 1992, \apj, 394, 628

\bibitem[{{Hartigan} {et~al.}(1995){Hartigan}, {Edwards}, \&
  {Ghandour}}]{HEG95}
{Hartigan}, P., {Edwards}, S., \& {Ghandour}, L. 1995, \apj, 452, 736

\bibitem[{{Hayashi}(1981)}]{HAYASHI81}
{Hayashi}, C. 1981, Progress of Theoretical Physics, 70, 35

\bibitem[{{Herbig} \& {Bell}(1988)}]{HB88}
{Herbig}, G.~H. \& {Bell}, K.~R. 1988, {Catalog of emission line stars of the
  orion population : 3 : 1988} (Lick Observatory Bulletin, Santa Cruz: Lick
  Observatory, |c1988)

\bibitem[{{Herbst} {et~al.}(1994){Herbst}, {Herbst}, {Grossman}, \&
  {Weinstein}}]{HERBST+94}
{Herbst}, W., {Herbst}, D.~K., {Grossman}, E.~J., \& {Weinstein}, D. 1994, \aj,
  108, 1906

\bibitem[{{Isella} \& {Natta}(2005)}]{IN05}
{Isella}, A. \& {Natta}, A. 2005, \aap, 438, 899

\bibitem[Isella et al.(2006)]{ITN06} Isella, A., Testi, L., 
\& Natta, A.\ 2006, \aap, 451, 951 

\bibitem[{{Johns-Krull}(2007)}]{JOHNS-KRULL07}
{Johns-Krull}, C.~M. 2007, ArXiv e-prints, 0704.2923

\bibitem[{{Kenyon} \& {Hartmann}(1995)}]{KH95}
{Kenyon}, S.~J. \& {Hartmann}, L. 1995, \apjs, 101, 117

\bibitem[{{Kholopov}(1959)}]{KHOLOPOV59}
{Kholopov}, P.~N. 1959, Soviet Astronomy, 3, 291

\bibitem[{{Koerner} \& {Sargent}(1995)}]{KS95}
{Koerner}, D.~W. \& {Sargent}, A.~I. 1995, \aj, 109, 2138

\bibitem[{{K{\"o}hler} {et~al.}(2000){K{\"o}hler}, {Kunkel}, {Leinert}, \&
  {Zinnecker}}]{KOHLER+00}
{K{\"o}hler}, R., {Kunkel}, M., {Leinert}, C., \& {Zinnecker}, H. 2000, \aap,
  356, 541

\bibitem[{{K\"{o}nigl}(1991)}]{KONIGL91}
{K\"{o}nigl}, A. 1991, \apjl, 370, L39

\bibitem[{{Leinert} {et~al.}(2004){Leinert}, {van Boekel}, {Waters},
  {Chesneau}, {Malbet}, {K{\" o}hler}, {Jaffe}, {Ratzka}, {Dutrey},
  {Preibisch}, {Graser}, {Bakker}, {Chagnon}, {Cotton}, {Dominik}, {Dullemond},
  {Glazenborg-Kluttig}, {Glindemann}, {Henning}, {Hofmann}, {de Jong},
  {Lenzen}, {Ligori}, {Lopez}, {Meisner}, {Morel}, {Paresce}, {Pel},
  {Percheron}, {Perrin}, {Przygodda}, {Richichi}, {Sch{\" o}ller}, {Schuller},
  {Stecklum}, {van den Ancker}, {von der L{\" u}he}, \& {Weigelt}}]{LEINERT+04}
{Leinert}, C., et al. 2004, \aap, 423, 537

\bibitem[{{Lin} {et~al.}(1996){Lin}, {Bodenheimer}, \& {Richardson}}]{LBR96}
{Lin}, D.~N.~C., {Bodenheimer}, P., \& {Richardson}, D.~C. 1996, \nat, 380, 606

\bibitem[{{Liu} {et~al.}(2007){Liu}, {Hinz}, {Meyer}, {Mamajek}, {Hoffmann},
  {Brusa}, {Miller}, \& {Kenworthy}}]{LIU+07}
{Liu}, W.~M., {Hinz}, P.~M., {Meyer}, M.~R., {Mamajek}, E.~E., {Hoffmann},
  W.~F., {Brusa}, G., {Miller}, D., \& {Kenworthy}, M.~A. 2007, \apj, 658, 1164

\bibitem[{{Marcy} {et~al.}(2005){Marcy}, {Butler}, {Fischer}, {Vogt}, {Wright},
  {Tinney}, \& {Jones}}]{MARCY+05}
{Marcy}, G., {Butler}, R.~P., {Fischer}, D., {Vogt}, S., {Wright}, J.~T.,
  {Tinney}, C.~G., \& {Jones}, H.~R.~A. 2005, Progress of Theoretical Physics
  Supplement, 158, 24

\bibitem[{{McCaughrean} \& {O'Dell}(1996)}]{MO96}
{McCaughrean}, M.~J. \& {O'Dell}, C.~R. 1996, \aj, 111, 1977

\bibitem[{{McLean} {et~al.}(2003){McLean}, {McGovern}, {Burgasser},
  {Kirkpatrick}, {Prato}, \& {Kim}}]{MCLEAN+03}
{McLean}, I.~S., {McGovern}, M.~R., {Burgasser}, A.~J., {Kirkpatrick}, J.~D.,
  {Prato}, L., \& {Kim}, S.~S. 2003, \apj, 596, 561

\bibitem[{{Millan-Gabet} {et~al.}(2007){Millan-Gabet}, {Malbet}, {Akeson},
  {Leinert}, {Monnier}, \& {Waters}}]{MILLAN-GABET+07}
{Millan-Gabet}, R., {Malbet}, F., {Akeson}, R., {Leinert}, C., {Monnier}, J.,
  \& {Waters}, R. 2007, in Protostars and Planets V, ed. B.~{Reipurth},
  D.~{Jewitt}, \& K.~{Keil}, 539--554

\bibitem[{{Millan-Gabet} {et~al.}(2001){Millan-Gabet}, {Schloerb}, \&
  {Traub}}]{MST01}
{Millan-Gabet}, R., {Schloerb}, F.~P., \& {Traub}, W.~A. 2001, \apj, 546, 358

\bibitem[{{Millan-Gabet} {et~al.}(1999){Millan-Gabet}, {Schloerb}, {Traub},
  {Malbet}, {Berger}, \& {Bregman}}]{MILLAN+99}
{Millan-Gabet}, R., {Schloerb}, F.~P., {Traub}, W.~A., {Malbet}, F., {Berger},
  J.~P., \& {Bregman}, J.~D. 1999, \apjl, 513, L131

\bibitem[{{Monin} {et~al.}(1998){Monin}, {Menard}, \& {Duchene}}]{MMD98}
{Monin}, J.-L., {Menard}, F., \& {Duchene}, G. 1998, \aap, 339, 113

\bibitem[{{Monnier} {et~al.}(2006){Monnier}, {Berger}, {Millan-Gabet}, {Traub},
  {Schloerb}, {Pedretti}, {Benisty}, {Carleton}, {Haguenauer}, {Kern},
  {Labeye}, {Lacasse}, {Malbet}, {Perraut}, {Pearlman}, \& {Zhao}}]{MONNIER+06}
{Monnier}, J.~D., et al. 2006, \apj, 647, 444

\bibitem[{{Monnier} \& {Millan-Gabet}(2002)}]{MM02}
{Monnier}, J.~D. \& {Millan-Gabet}, R. 2002, \apj, 579, 694

\bibitem[{{Monnier} {et~al.}(2005){Monnier}, {Millan-Gabet}, {Billmeier},
  {Akeson}, {Wallace}, {Berger}, {Calvet}, {D'Alessio}, {Danchi}, {Hartmann},
  {Hillenbrand}, {Kuchner}, {Rajagopal}, {Traub}, {Tuthill}, {Boden}, {Booth},
  {Colavita}, {Gathright}, {Hrynevych}, {Le Mignant}, {Ligon}, {Neyman},
  {Swain}, {Thompson}, {Vasisht}, {Wizinowich}, {Beichman}, {Beletic},
  {Creech-Eakman}, {Koresko}, {Sargent}, {Shao}, \& {van Belle}}]{MONNIER+05}
{Monnier}, J.~D., et al. 2005, \apj, 624, 832

\bibitem[{{Mora} {et~al.}(2001){Mora}, {Mer{\'{\i}}n}, {Solano}, {Montesinos},
  {de Winter}, {Eiroa}, {Ferlet}, {Grady}, {Davies}, {Miranda}, {Oudmaijer},
  {Palacios}, {Quirrenbach}, {Harris}, {Rauer}, {Cameron}, {Deeg}, {Garz{\'
  o}n}, {Penny}, {Schneider}, {Tsapras}, \& {Wesselius}}]{MORA+01}
{Mora}, A., et al. 2001, \aap, 378, 116

\bibitem[{{Muzerolle} {et~al.}(2001){Muzerolle}, {Calvet}, \&
  {Hartmann}}]{MCH01}
{Muzerolle}, J., {Calvet}, N., \& {Hartmann}, L. 2001, \apj, 550, 944

\bibitem[{{Muzerolle} {et~al.}(1998){Muzerolle}, {Hartmann}, \&
  {Calvet}}]{MHC98}
{Muzerolle}, J., {Hartmann}, L., \& {Calvet}, N. 1998, \aj, 116, 2965

\bibitem[{{Perryman} {et~al.}(1997){Perryman}, {Lindegren}, {Kovalevsky},
  {Hoeg}, {Bastian}, {Bernacca}, {Cr{\'e}z{\'e}}, {Donati}, {Grenon}, {van
  Leeuwen}, {van der Marel}, {Mignard}, {Murray}, {Le Poole}, {Schrijver},
  {Turon}, {Arenou}, {Froeschl{\'e}}, \& {Petersen}}]{PERRYMAN+97}
{Perryman}, M.~A.~C., et al. 1997, \aap, 323, L49

\bibitem[{{Pollack} {et~al.}(1994){Pollack}, {Hollenbach}, {Beckwith},
  {Simonelli}, {Roush}, \& {Fong}}]{POLLACK+94}
{Pollack}, J.~B., {Hollenbach}, D., {Beckwith}, S., {Simonelli}, D.~P.,
  {Roush}, T., \& {Fong}, W. 1994, \apj, 421, 615

\bibitem[{{Prato} {et~al.}(2003){Prato}, {Greene}, \& {Simon}}]{PGS03}
{Prato}, L., {Greene}, T.~P., \& {Simon}, M. 2003, \apj, 584, 853

\bibitem[{{Shu} {et~al.}(1994){Shu}, {Najita}, {Ostriker}, {Wilkin}, {Ruden},
  \& {Lizano}}]{SHU+94}
{Shu}, F., {Najita}, J., {Ostriker}, E., {Wilkin}, F., {Ruden}, S., \&
  {Lizano}, S. 1994, \apj, 429, 781

\bibitem[{{Simon} {et~al.}(1995){Simon}, {Ghez}, {Leinert}, {Cassar}, {Chen},
  {Howell}, {Jameson}, {Matthews}, {Neugebauer}, \& {Richichi}}]{SIMON+95}
{Simon}, M., {Ghez}, A.~M., {Leinert}, C., {Cassar}, L., {Chen}, W.~P.,
  {Howell}, R.~R., {Jameson}, R.~F., {Matthews}, K., {Neugebauer}, G., \&
  {Richichi}, A. 1995, \apj, 443, 625

\bibitem[{{Skrutskie} {et~al.}(1996){Skrutskie}, {Meyer}, {Whalen}, \&
  {Hamilton}}]{SKRUTSKIE+96}
{Skrutskie}, M.~F., {Meyer}, M.~R., {Whalen}, D., \& {Hamilton}, C. 1996, \aj,
  112, 2168

\bibitem[{{Steenman} \& {Th\'{e}}(1991)}]{ST91}
{Steenman}, H. \& {Th\'{e}}, P.~S. 1991, \apss, 184, 9

\bibitem[Tannirkulam et al.(2007)]{THM07} Tannirkulam, A., 
Harries, T.~J., \& Monnier, J.~D.\ 2007, \apj, 661, 374 

\bibitem[{{Valenti} {et~al.}(1993){Valenti}, {Basri}, \& {Johns}}]{VBJ93}
{Valenti}, J.~A., {Basri}, G., \& {Johns}, C.~M. 1993, \aj, 106, 2024

\bibitem[{{van Boekel} {et~al.}(2004){van Boekel}, {Min}, {Leinert}, {Waters},
  {Richichi}, {Chesneau}, {Dominik}, {Jaffe}, {Dutrey}, {Graser}, {Henning},
  {de Jong}, {K{\" o}hler}, {de Koter}, {Lopez}, {Malbet}, {Morel}, {Paresce},
  {Perrin}, {Preibisch}, {Przygodda}, {Sch{\" o}ller}, \&
  {Wittkowski}}]{VANBOEKEL+04}
{van Boekel}, R., et al. 2004, \nat, 432, 479

\bibitem[{{Vinkovi{\'c}} {et~al.}(2006){Vinkovi{\'c}}, {Ivezi{\'c}},
  {Jurki{\'c}}, \& {Elitzur}}]{VINKOVIC+06}
{Vinkovi{\'c}}, D., {Ivezi{\'c}}, {\v Z}., {Jurki{\'c}}, T., \& {Elitzur}, M.
  2006, \apj, 636, 348

\bibitem[{{Vinkovi{\'c}} \& {Jurki{\'c}}(2007)}]{VJ07}
{Vinkovi{\'c}}, D. \& {Jurki{\'c}}, T. 2007, \apj, 658, 462

\bibitem[{{Walker}(1972)}]{WALKER72}
{Walker}, M.~F. 1972, \apj, 175, 89

\bibitem[{{Walter} {et~al.}(1994){Walter}, {Vrba}, {Mathieu}, {Brown}, \&
  {Myers}}]{WALTER+94}
{Walter}, F.~M., {Vrba}, F.~J., {Mathieu}, R.~D., {Brown}, A., \& {Myers},
  P.~C. 1994, \aj, 107, 692

\bibitem[{{White} \& {Ghez}(2001)}]{WG01}
{White}, R.~J. \& {Ghez}, A.~M. 2001, \apj, 556, 265

\bibitem[{{Wichmann} {et~al.}(2000){Wichmann}, {Torres}, {Melo}, {Frink},
  {Allain}, {Bouvier}, {Krautter}, {Covino}, \& {Neuh{\"a}user}}]{WICHMANN+00}
{Wichmann}, R., {Torres}, G., {Melo}, C.~H.~F., {Frink}, S., {Allain}, S.,
  {Bouvier}, J., {Krautter}, J., {Covino}, E., \& {Neuh{\"a}user}, R. 2000,
  \aap, 359, 181

\bibitem[{{Zuckerman} \& {Song}(2004)}]{ZS04}
{Zuckerman}, B. \& {Song}, I. 2004, \araa, 42, 685

\end{thebibliography}

\begin{deluxetable}{lccccccc}
\tabletypesize{\tiny}
\tablewidth{0pt}
\tablecaption{Target and Calibrator Properties and Observations 
\label{tab:sample}}
\tablehead{\colhead{Source} & \colhead{$\alpha$} 
& \colhead{$\delta$} & \colhead{$d$} & \colhead{Type} & \colhead{KI}
& \colhead{NIRSPEC} & \colhead{References}}
\startdata
\multicolumn{8}{c}{Target Stars} \\
\hline
CI Tau & 04 33 52.01 & +22 50 30.1 & 140 & K7 & 24 Oct 2005 & 07 Jan 2006 & 
1 \\
DK Tau A & 04 30 44.28 & +26 01 24.6 & 140 & K9 & 30 Oct 2004, 24 Oct 2005 & 
19 Nov 2004, 07 Jan 2006 & 2 \\
DK Tau B & 04 30 44.42 & +26 01 23.9 & 140 & M1 & 30 Oct 2004, 24 Oct 2005 & 
19 Nov 2004, 07 Jan 2006 & 2 \\
AA Tau & 04 34 55.45 & +24 28 53.7 & 140 & K7 & 30 Oct 2004, 24 Oct 2005 & 
19 Nov 2004, 07 Jan 2006 & 1 \\
RW Aur A & 05 07 49.568 & +30 24 05.161 & 140 & K1 & 30 Oct 2004$^{\ast}$ & 
20 Nov 2004, 07 Jan 2006 & 3 \\
RW Aur B & 05 07 49.465 & +30 24 04.82 & 140 & K5 & 30 Oct 2004 & 
20 Nov 2004, 07 Jan 2006 & 3 \\
V1002 Sco & 16 12 40.505 & -18 59 28.14 & 160 & K0 & 
21 Apr 2005 & 01 Jun 2005 & 4 \\
AS 206 & 16 25 56.3 & -24 20 50.0 & 160 & K6.5 & 21 Apr 2005 & 13 Jul 2006 & 
5\\
V1331 Cyg & 21 01 09.21 & +50 21 44.8 & 700 & G5 & 24 Oct 2005 & 13 Jul 2006 &
6,7 \\
DI Cep & 22 56 11.534 & +58 40 01.795 & 300 & G8 & 24 Oct 2005 & 13 Jul 2006 & 
5,8 \\
BM And & 23 37 38.47 & +48 24 12.0 & 440 & K5 & 
24 Oct 2005 & 07 Jan 2006 & 9,10 \\
\hline
\multicolumn{8}{c}{KI Calibrator Stars} \\
\hline
HD 28459 & 04 30 38.408 & +32 27 27.965 & 94 & B9.5V & 
30 Oct 2004, 24 Oct 2005 & & 11 \\
HD 282230 & 04 31 23.148 & +31 16 41.390 & 77 & K0V & 
30 Oct 2004, 24 Oct 2005 & & 11 \\
HD 30378 & 04 48 22.743 & +29 46 22.811 & 196 & B9.5V & 
30 Oct 2004, 24 Oct 2005 & & 11 \\
HD 36724 & 05 34 58.506 & +26 58 19.958 & 86 & F7V & 
30 Oct 2004, 24 Oct 2005 & & 11 \\
HDC142943 & 15 57 57.772 & -17 05 23.974 & 75 & F6V & 21 Apr 2005 & & 11 \\
HDC144925 & 16 09 02.600 & -18 59 44.037 & 150 & A0V & 21 Apr 2005 & & 11\\
HD 214279 & 22 35 51.994 & +56 04 14.067 & 108 & A3V & 24 Oct 2005 & & 11\\
HD 214240 & 22 35 53.382 & +50 04 14.836 & 500 & B3V & 24 Oct 2005 & & 11 \\
HD 219290 & 23 14 14.385 & +50 37 04.408 & 126 & A0V & 24 Oct 2005 & & 11 \\
HD 219891 & 23 19 02.413 & +45 08 12.302 & 126 & A5V & 24 Oct 2005 & & 11 \\
\hline
\multicolumn{8}{c}{NIRSPEC Calibrator Stars} \\
\hline
HR 1251 & 04 03 09.3801 & +05 59 21.498 & 40 & A1V & & 
19 Nov 2004, 07 Jan 2006 & 11 \\
HR 1252 & 04 59 15.4096 & +37 53 24.881 & 49 & A1V & & 
20 Nov 2004 & 11 \\
HR 5511 & 14 46 14.9241 & +01 53 34.388 & 39 & A0V & & 13 Jul 2006 & 11 \\ 
HR 8518 & 22 21 39.3754 & -01 23 14.393 & 48 & A0V & & 01 Jun 2005 & 11 \\
\hline
\multicolumn{8}{c}{NIRSPEC WLTT Template Stars} \\
\hline
AG Tri A & 02 27 29.2543 & +30 58 24.613 & 42 & K6 & & 
19, 20 Nov 2004, 07 Jan 2006, 13 Jul 2006 & 11 \\
AG Tri B & 02 27 29.2543 & +30 58 24.613 & 42 & M0 & & 
19, 20 Nov 2004, 07 Jan 2006, 13 Jul 2006 & 11 \\
TYC 5882 & 04 02 16.5 & -15 21 30 & 63 & K3/4  & & 20 Nov 2004 & 12 \\ 
RX J0409.8 & 04 09 51.108 & +24 46 21.20 & 140 & M1.5 & & 
19, 20 Nov 2004, 07 Jan 2006 & 13 \\
RX J0432.8 & 04 32 53.237 & +17 35 33.65 & 140 & M2 & & 
19, 20 Nov 2004, 07 Jan 2006 & 13 \\
RX J0437.4A & 04 37 26.89 & +18 51 25.9 & 140 & K6 & & 
19, 20 Nov 2004, 07 Jan 2006 & 13 \\
RX J0437.4B & 04 37 26.89 & +18 51 25.9 & 140 & M0.5 & & 
19, 20 Nov 2004, 07 Jan 2006 & 13 \\
RX J0438.2 & 04 38 15.619 & +23 02 27.88 & 140 & M2 & & 
19, 20 Nov 2004, 07 Jan 2006 & 13 \\
RX J1534.3 &  15 34 23.1 & -33 00 07 & 145 & M0 &  & 13 Jul 2006 & 14 \\ 
RX J1546.0 & 15 46 05.5 & -29 20 40 & 145 & M0 &  & 
01 Jun 2005, 13 Jul 2006 & 14 \\
RX J1546.7 & 15 46 47.0 & -32 10 06 & 145 & M2 & & 01 Jun 2005 & 14   \\
RX J1557.8 & 15 57 50.0 & -23 05 09 & 145 & M0 & & 01 Jun 2005 & 14 \\
RX J1558.5 & 15 58 53.6 & -25 12 32  & 145 & M1 & & 01 Jun 2005 & 14  \\
\enddata
\tablecomments{$\ast$--for RW Aur A, we also utilize KI measurements from
23 Oct 2002 \citep{AKESON+05b}. All Taurus sources were assumed to 
be at a distance of 140 pc.}  
\tablerefs{(1) \cite{KH95}; (2) \cite{MMD98}; (3) \cite{WG01}; (4) 
\cite{WALTER+94}; (5) \cite{HB88}; (6) \cite{CK81}; (7) \cite{HP92}; 
(8) \cite{KHOLOPOV59}; (9) \cite{AH69}; (10) \cite{MORA+01}; 
(11) \cite{PERRYMAN+97}; (12) \cite{ZS04}; (13) \cite{WICHMANN+00}; 
(14) \cite{KOHLER+00}}
\end{deluxetable}

\begin{deluxetable}{lcccc}
\tabletypesize{\footnotesize}
\tablewidth{0pt}
\tablecaption{PAIRITEL $JHK_{\rm s}$ Photometry \label{tab:phot}}
\tablehead{\colhead{Source} & \colhead{Date} 
& \colhead{$m_J$} & \colhead{$m_H$} & \colhead{$m_{K_{\rm s}}$}}
\startdata
CI Tau & 24Oct05 & $8.50 \pm 0.02$ & $7.99 \pm 0.02$ & $7.51 \pm 0.03$ \\
 & 20Nov05 &  $8.25 \pm 0.02$  & $7.83 \pm 0.02$ & $7.34 \pm 0.02$ \\
DK Tau AB & 18Oct04 & $7.94 \pm 0.02$ & $7.46 \pm 0.02$ & $7.03 \pm 0.03$ \\
 & 05Nov04 &  $7.95 \pm 0.02$  & $7.45 \pm 0.02$ & $6.98 \pm 0.02$ \\
 & 24Oct05 &  $7.83 \pm 0.05$  & $7.38 \pm 0.05$ & $6.92 \pm 0.08$ \\
AA Tau & 18Oct04 & $8.80 \pm 0.02$  & $8.37 \pm 0.02$ & $7.94 \pm 0.03$ \\
 & 05Nov04 &  $9.10 \pm 0.03$  & $8.58 \pm 0.03$ & $8.12 \pm 0.03$ \\
 & 24Oct05 &  $8.82 \pm 0.03$  & $8.38 \pm 0.03$ & $7.74 \pm 0.04$ \\
 & 20Nov05 &  $8.65 \pm 0.03$  & $8.20 \pm 0.03$ & $7.82 \pm 0.03$ \\
RW Aur AB & 18Oct04 & $8.63 \pm 0.02$ & $8.19 \pm 0.02$ & $7.68 \pm 0.02$ \\
 & 24Oct05 & $8.51 \pm 0.02$  & $8.00 \pm 0.02$ & $7.54 \pm 0.02$ \\
V1002 Sco & 21Apr05 & $8.29 \pm 0.02$ & $7.87 \pm 0.02$ & $7.35 \pm 0.02$ \\
AS 206$^{\ast}$ & 21Apr05 & $9.0 \pm 0.3$ & $8.5 \pm 0.3$ & $7.9 \pm 0.3$ \\ 
V1331 Cyg$^{\ast}$ & 12Sep04 & $10.01 \pm 0.04$ & $9.64 \pm 0.04$ & 
$8.98 \pm 0.04$ \\
 & 13Oct04 & $10.02 \pm 0.04$ & $9.65 \pm 0.04$ & $9.06 \pm 0.04$ \\
 & 23Oct04 & $10.03 \pm 0.04$ & $9.57 \pm 0.04$ & $9.00 \pm 0.04$ \\
 & 17Nov05 & $10.06 \pm 0.04$ & $9.63 \pm 0.04$ & $9.03 \pm 0.04$ \\
 & 24Oct05 & $10.06 \pm 0.04$ & $9.67 \pm 0.04$ & $9.07 \pm 0.04$ \\
 & 25Oct05 & $10.01 \pm 0.04$ & $9.64 \pm 0.04$ & $9.04 \pm 0.04$ \\
 & 26Oct05 & $9.93 \pm 0.04$ & $9.54 \pm 0.04$ & $8.95 \pm 0.04$ \\
DI Cep$^{\ast}$ & 13Oct04 & $9.1 \pm 0.2$ & $8.6 \pm 0.2$ & $8.2 \pm 0.2$ \\
 & 20Oct04 & $9.2 \pm 0.2$ & $8.8 \pm 0.2$ & $8.3 \pm 0.2$ \\ 
BM And & 13Oct04 & $9.85 \pm 0.03$ & $9.38 \pm 0.03$ & $8.89 \pm 0.03$ \\
 & 23Oct04 & $10.12 \pm 0.03$ & $9.59 \pm 0.03$ & $9.14 \pm 0.03$ \\ 
 & 24Oct05 & $9.61 \pm 0.02$ & $9.10 \pm 0.02$ & $8.61 \pm 0.02$ \\ 
 & 17Nov05 & $9.84 \pm 0.03$ & $9.32 \pm 0.03$ & $8.85 \pm 0.03$ \\ 
\enddata
\tablecomments{$\ast$--For AS 206, V1331 Cyg, and DI Cep, our photometric 
calibrators appear to be variable at a low level, and the quoted uncertainties
include this potential error.}
\end{deluxetable}

\begin{deluxetable}{lcccc}
\tabletypesize{\footnotesize}
\tablewidth{0pt}
\tablecaption{$K$-band veiling measurements \label{tab:veilings}}
\tablehead{\colhead{Source} & \colhead{Date} 
& \colhead{$r_K$} & \colhead{Rotational Broadening (km s$^{-1}$)} & 
\colhead{Template Star}}
\startdata
CI Tau & 07 Jan 2006 & $1.16 \pm 0.13$ & $<12$ & RXJ 0437.4A \\
DK Tau A & 19 Nov 2004 & $1.58 \pm 0.16$ & $<12$ & RX J0437.4B \\ 
 & 07 Jan 2006 & $1.32 \pm 0.14$ & $<12$ & RX J0437.4B \\
DK Tau B & 19 Nov 2004 & $0.78 \pm 0.10$ & $<12$ & RX J0437.4B \\ 
 & 07 Jan 2006 & $0.80 \pm 0.15$ & $<12$ & RX J0437.4B \\
AA Tau & 19 Nov 2004 & $0.54 \pm 0.06$ & $<12$ &  RX J0437.4B \\ 
 & 07 Jan 2006 & $0.34 \pm 0.06$ & $<12$ &  RX J0437.4B \\
RW Aur A & 20 Nov 2004 & $3.26^{+0.86}_{-0.62}$ & $35 \pm 10$ & 
AG Tri A \\
& 07 Jan 2006 & $1.50 \pm 0.20$ & $35 \pm 10$ & AG Tri A \\
RW Aur B & 20 Nov 2004 & $0.32 \pm 0.08$ & $11 \pm 4$ &  RX J0437.4B  \\
& 07 Jan 2006 & $0.28 \pm 0.06$ & $11 \pm 4$ &  RX J0437.4B \\
V1002 Sco & 01 Jun 2005 & $0.36 \pm 0.08$ & $78 \pm 6$ & RX J1557.8 \\ 
AS 206 & 13 Jul 2006 & $1.48 \pm 0.22$ & $8 \pm 7$ & AG Tri A \\ 
V1331 Cyg & 13 Jul 2006 & $>10$ \\
DI Cep & 13 Jul 2006 & $3.40 \pm 0.71$ & $9 \pm 5$ & AG Tri A \\
BM And & 07 Jan 2006 & $2.30^{+0.44}_{-1.50}$ & $17 \pm 17$ & TYC 5882 \\ 
\enddata
\tablecomments{The $r_K$ and rotational broadening values were
determined by minimizing the $\chi^2$ residuals between observed spectra
and a grid of veiled, rotationally broadened template spectra 
(\S \ref{sec:rk}).  The template stars are WLTTs showing no near-IR
excess emission, and fitted veilings for these objects are consistent
with zero.  However, the templates may have non-negligible $v\sin i$ 
values, and thus the fitted rotational broadenings 
for our targets are lower limits for the true $v \sin i$ values.  
Quoted error bars are 1$\sigma$ statistical
uncertainties in the best-fit parameters. 
Spectral types of template stars can be found in 
Table \ref{tab:sample}.
}
\end{deluxetable}

\begin{deluxetable}{lccc}
\tabletypesize{\footnotesize}
\tablewidth{0pt}
\tablecaption{Face-on Ring Radii \label{tab:uds}}
\tablehead{\colhead{Source} & \colhead{Date} 
& \colhead{$R_{\rm ring}$ (AU)} & \colhead{$R_{\rm ring}$ (fixed $r_K$)}}
\startdata
CI Tau & 24Oct05 & 0.097 $\pm$ 0.008 & 0.097 $\pm$ 0.008 \\
DK Tau A & 30Oct04 & $0.103 \pm 0.005$ & 0.105 $\pm$ 0.005 \\
 & 24Oct05 & $0.107 \pm 0.008$ & 0.105 $\pm$ 0.008 \\
DK Tau B & 30Oct04 & $0.137 \pm 0.018$ & 0.137 $\pm$ 0.012 \\
 & 24Oct05 & $0.137 \pm 0.018$ & 0.138 $\pm$ 0.018 \\
AA Tau & 30Oct04 & $0.084 \pm 0.029$ & 0.090 $\pm$ 0.031 \\
 & 24Oct05 & $0.133 \pm 0.019$ &  0.120 $\pm$ 0.016 \\
RW Aur A & 23Oct02 & 0.103 $\pm$ 0.005 & $0.111 \pm 0.007$ \\
 & 30Oct04 & 0.181 $\pm$ 0.016 & $0.131 \pm 0.007$ \\
RW Aur B & 24Oct05 & 0.161 $\pm$ 0.034  & 0.161 $\pm$ 0.034 \\
V1002 Sco & 21Apr05 & 0.118 $\pm$ 0.027 & 0.118 $\pm$ 0.027 \\
AS 206 & 21Apr05 & 0.112 $\pm$ 0.007 & 0.112 $\pm$ 0.007 \\
V1331 Cyg & 24Oct05 & 0.315 $\pm$ 0.021  & 0.315 $\pm$ 0.021 \\
DI Cep & 24Oct05 & 0.165 $\pm$ 0.039 & 0.165 $\pm$ 0.039 \\
BM And & 24Oct05 & 0.249 $\pm$ 0.028  & 0.249 $\pm$ 0.028 \\
\enddata
\tablecomments{$R_{\rm ring}$ is computed from fits of uniform ring models
to KI data, and the distances assumed in Table \ref{tab:sample}.  Quoted
error bars are 1$\sigma$ statistical uncertainties in the $\chi^2$ minimzation
between models and measured $V_{\rm disk}^2$. Column 3 lists the ring radii
determined when $V^2_{\rm disk}$ are computed using the veiling measurement
closest in time to the KI measurement.  Column 4 lists the ring radii for 
$V^2_{\rm disk}$ estimated  using the average veiling measured over all epochs.
}
\end{deluxetable}

\begin{deluxetable}{lccccc}
\tabletypesize{\footnotesize}
\tablewidth{0pt}
\tablecaption{Br$\gamma$ Emission Properties \label{tab:ews}}
\tablehead{\colhead{Source} & \colhead{Date} 
& \colhead{EW (\AA)} & \colhead{$L_{\rm Br \gamma}$ 
($10^{-4}$ L$_{\odot}$)} & \colhead{$L_{\rm acc}$ (L$_{\odot}$)} & 
\colhead{$L_{\ast}$ (L$_{\odot}$)}}
\startdata
CI Tau & 07 Jan 2006 & $7.4 \pm 0.1$ & 2.3 & 0.70 & 0.8 \\
DK Tau A & 19 Nov 2004 & $7.6 \pm 0.1$ & 2.6 & 0.80 & 1.7 \\
 & 07 Jan 2006 & $6.9 \pm 0.1$ & 2.5 & 0.76 & 1.7 \\
DK Tau B & 19 Nov 2004 & $2.7 \pm 0.1$ & 0.3 & 0.05 & 0.5 \\
 & 07 Jan 2006 & $1.8 \pm 0.1$ & 0.2 & 0.03 & 0.5 \\
AA Tau & 19 Nov 2004 & $4.8 \pm 0.1$ & 0.7 & 0.16 & 0.6 \\
 & 07 Jan 2006 & $0.2 \pm 0.1$ & $<0.1$ & $<0.01$ & 0.6 \\
RW Aur A & 20 Nov 2004 & $9.3 \pm 0.1$ & 1.5 & 0.42 & 1.7 \\
 & 07 Jan 2006 & $16.5 \pm 0.1$ & 3.1 & 1.02 & 1.7 \\
RW Aur B & 20 Nov 2004 & $6.7 \pm 0.1$ & 0.3 & 0.05 & 0.4 \\
 & 07 Jan 2006 & $5.1 \pm 0.1$ & 0.3 & 0.04 & 0.4 \\
V1002 Sco & 01 Jun 2005 & $14 \pm 6$ & 5.4 & 2.07 & 3.8 \\
AS 206 & 13 Jul 2006 & $9.7 \pm 0.2$ & 2.4 & 0.74 & 1.6 \\
V1331 Cyg & 13 Jul 2006 & $14.4 \pm 0.2$ & 24.5 & 13.78 & 21.0 \\
DI Cep & 13 Jul 2006 & $8.4 \pm 0.2$ & 4.6 & 1.69 & 5.2 \\
BM And & 07 Jan 2006 & $1.4 \pm 0.1$ & 1.0 & 0.23 & 5.5 \\
\enddata
\tablecomments{Given the uncertainties in the photometry and EWs, we estimate
uncertainties between 5\% and 20\% for $L_{\rm Br \gamma}$.  The conversion
from $L_{\rm Br \gamma}$ to $L_{\rm acc}$ includes additional uncertainties
\citep{MHC98,MCH01}, and $L_{\rm acc}$ is thus uncertain by a factor
of a few. We also list stellar luminosities, $L_{\ast}$,  drawn
from the literature \citep{WG01,HEG95,WALTER+94,HP92}.}
\end{deluxetable}

\clearpage

\epsscale{1.0}

\epsscale{1.0}

\epsscale{1.0}

\epsscale{1.0}

\epsscale{0.9}

\epsscale{1.0}

\end{document}